\documentclass[trackchanges,twocolumn]{aastex701}

\usepackage{graphicx}
\usepackage{aas_macros}
\usepackage{xspace}
\usepackage{longtable}
\newcommand{\tdes}{\texttt{tdescore}\xspace}
\newcommand{\name}{TDE 2025abcr\xspace}
\newcommand{\host}{WISEA J014656.04-152214.7\xspace}
\newcommand{\xmm}{3XMM J2150\xspace}
\newcommand{\erass}{eRASS J142140.3-295325\xspace}
\newcommand{\wings}{WINGS J134849.88+263557.5\xspace}
\newcommand{\Msol}{\mbox{$\mathrm{M}_{\odot}$}\xspace}
\newcommand{\Mstar}{\mbox{$\mathrm{M}_{\star}$}\xspace}
\newcommand{\galaxymass}{$10^{11.18 \pm 0.03}$\Msol}
\newcommand{\bhmass}{$10^{8.82 \pm 0.65}$\Msol}
\newcommand{\tdemass}{$10^{6.09\pm0.53}$\Msol}
\newcommand{\tpeak}{2025-11-14\xspace}

\newcommand{\sncosmo}{\texttt{sncosmo}\xspace}

\newcommand{\prospector}{\texttt{prospector}\xspace}
\newcommand{\uvotredux}{\texttt{uvotredux}\xspace}
\newcommand{\galsynthspec}{\texttt{galsynthspec}\xspace}

\newcommand{\scipy}{\texttt{scipy}\xspace}
\newcommand{\swift}{\textit{Neil Gehrels Swift Observatory}\xspace}

\begin{document}

\title{\name: A Tidal Disruption Event in the Outskirts of a Massive Galaxy}

\author[orcid=0000-0003-2434-0387]{Robert Stein}
\email[show]{rdstein@umd.edu}
\altaffiliation{Neil Gehrels Prize Postdoctoral Fellow}
\affiliation{Department of Astronomy, University of Maryland, College Park, MD 20742, USA}
\affiliation{Joint Space-Science Institute, University of Maryland, College Park, MD 20742, USA} 
\affiliation{Astrophysics Science Division, NASA Goddard Space Flight Center, Mail Code 661, Greenbelt, MD 20771, USA} 

\author[0000-0001-8544-584X]{Jonathan Carney}
\email{jcarney@unc.edu}
\affiliation{Department of Physics and Astronomy, University of North Carolina at Chapel Hill, Chapel Hill, NC 27599-3255, USA}

\author[0000-0002-4557-6682]{Charlotte Ward}
\email{cvw5890@psu.edu}
\affiliation{Department of Astronomy \& Astrophysics, 525 Davey Lab, 251 Pollock Road, The Pennsylvania State University, University Park, PA 16802, USA}

\author[0000-0003-4768-7586]{Raffaella Margutti}
\email{rmargutti@berkeley.edu}
\affiliation{Department of Astronomy, University of California, Berkeley, CA 94720-3411, USA}
\affiliation{Department of Physics, University of California, Berkeley, CA 94720-7300, USA} \affiliation{Berkeley Center for Multi-messenger Research on Astrophysical Transients and Outreach (Multi-RAPTOR), University of California, Berkeley, CA 94720-3411, USA}

\author[0000-0002-9364-5419]{Xander J. Hall}
\affiliation{McWilliams Center for Cosmology and Astrophysics, Department of Physics, Carnegie Mellon University, 5000 Forbes Avenue, Pittsburgh, PA 15213}
\email{xhall@cmu.edu}

\author[0000-0003-0466-3779]{Itai Sfaradi}
\email{Itai.sfaradi@berkeley.edu}
\affiliation{Department of Astronomy, University of California, Berkeley, CA 94720-3411, USA}
\affiliation{Berkeley Center for Multi-messenger Research on Astrophysical Transients and Outreach (Multi-RAPTOR), University of California, Berkeley, CA 94720-3411, USA}


\author[0000-0002-8977-1498]{Igor Andreoni}
\email{igor.andreoni@unc.edu}
\affiliation{Department of Physics and Astronomy, University of North Carolina at Chapel Hill, Chapel Hill, NC 27599-3255, USA}

\author[0000-0002-0326-6715]{Panos Charalampopoulos}
\email{panos.charalampopoulos.astro@gmail.com}
\affiliation{Finnish Centre for Astronomy with ESO (FINCA), FI-20014 University of Turku, Finland}

\author[0000-0002-7706-5668]{Ryan~Chornock}
\email{chornock@berkeley.edu}
\affiliation{Department of Astronomy, University of California, Berkeley, CA 94720-3411, USA}
\affiliation{Berkeley Center for Multi-messenger Research on Astrophysical Transients and Outreach (Multi-RAPTOR), University of California, Berkeley, CA 94720-3411, USA}

\author[0000-0003-3703-5154]{Suvi Gezari}
\email{suvi@umd.edu}
\affiliation{Department of Astronomy, University of Maryland, College Park, MD 20742, USA}

\author[0000-0001-6331-112X]{Geoffrey Mo}
\email{gmo@caltech.edu}
\affiliation{Division of Physics, Mathematics and Astronomy, California Institute of Technology, Pasadena, CA 91125, USA}
\affiliation{The Observatories of the Carnegie Institution for Science, Pasadena, CA 91101, USA}

\author[0000-0001-6747-8509]{Yuhan Yao}
\email{yuhanyao@berkeley.edu}
\affiliation{Miller Institute for Basic Research in Science, 206B Stanley Hall, Berkeley, CA 94720, USA}
\affiliation{Department of Astronomy, University of California, Berkeley, CA 94720-3411, USA}
\affiliation{Berkeley Center for Multi-messenger Research on Astrophysical Transients and Outreach (Multi-RAPTOR), University of California, Berkeley, CA 94720-3411, USA}


\author[0000-0002-8935-9882]{Akash Anumarlapudi}
\email{akasha@unc.edu}
\affiliation{Department of Physics and Astronomy, University of North Carolina at Chapel Hill, Chapel Hill, NC 27599-3255, USA}

\author[0000-0001-8018-5348]{Eric C. Bellm}\email{ecbellm@uw.edu}
\affiliation{DIRAC Institute, Department of Astronomy, University of Washington, 3910 15th Avenue NE, Seattle, WA 98195, USA}

\author[0000-0002-7777-216X]{Joshua S. Bloom}\email{joshbloom@berkeley.edu}
\affiliation{Department of Astronomy, University of California, Berkeley, CA 94720-3411, USA}

\author[0009-0001-0574-2332]{Malte Busmann}
\email{m.busmann@physik.lmu.de}
\affiliation{University Observatory, Faculty of Physics, Ludwig-Maximilians-Universität München, Scheinerstr. 1, 81679 Munich, Germany}
\affiliation{Excellence Cluster ORIGINS, Boltzmannstr. 2, 85748 Garching, Germany}

\author[orcid=0000-0002-4770-5388]{Ilaria Caiazzo} 
\email{ilaria.caiazzo@ist.ac.at}
\affiliation{Institute of Science and Technology Austria, Am Campus 1, 3400 Klosterneuburg, Austria}

\author[0000-0003-1673-970X]{S.~Bradley Cenko}
\email{brad.cenko@nasa.gov}
\affiliation{Astrophysics Science Division, NASA Goddard Space Flight Center, Mail Code 661, Greenbelt, MD 20771, USA}
\affiliation{Joint Space-Science Institute, University of Maryland, College Park, MD 20742, USA}

\author[0000-0002-3168-0139]{Matthew J. Graham}
\email{mjg@caltech.edu}
\affiliation{Division of Physics, Mathematics and Astronomy, California Institute of Technology, Pasadena, CA 91125, USA}

\author[0000-0001-5668-3507]{Steven L. Groom}\email{sgroom@ipac.caltech.edu}
\affiliation{IPAC, California Institute of Technology, 1200 E. California Blvd, Pasadena, CA 91125, USA}

\author[0000-0003-3270-7644]{Daniel Gruen}
\email{daniel.gruen@lmu.de}
\affiliation{University Observatory, Faculty of Physics, Ludwig-Maximilians-Universität München, Scheinerstr. 1, 81679 Munich, Germany}
\affiliation{Excellence Cluster ORIGINS, Boltzmannstr. 2, 85748 Garching, Germany}

\author[0000-0002-5698-8703]{Erica Hammerstein}
\email{ekhammer@berkeley.edu}
\affiliation{Department of Astronomy, University of California, Berkeley, CA 94720-3411, USA}
\affiliation{Berkeley Center for Multi-messenger Research on Astrophysical Transients and Outreach (Multi-RAPTOR), University of California, Berkeley, CA 94720-3411, USA}

\author[0000-0003-1970-4684]{Benjamin C. Kaiser}
\email{ben.kaiser@unc.edu}
\affiliation{Department of Physics and Astronomy, University of North Carolina at Chapel Hill, Chapel Hill, NC 27599-3255, USA}

\author[0000-0002-5619-4938]{Mansi M. Kasliwal}\email{mansi@astro.caltech.edu}
\affiliation{Division of Physics, Mathematics and Astronomy, California Institute of Technology, Pasadena, CA 91125, USA}

\author[0000-0002-9700-0036]{Brendan O'Connor}
\email{boconno2@andrew.cmu.edu}  
\affiliation{McWilliams Center for Cosmology and Astrophysics, Department of Physics, Carnegie Mellon University, 5000 Forbes Avenue, Pittsburgh, PA 15213}

\author[0000-0002-6011-0530]{Antonella Palmese}
\affiliation{McWilliams Center for Cosmology and Astrophysics, Department of Physics, Carnegie Mellon University, 5000 Forbes Avenue, Pittsburgh, PA 15213}
\email{palmese@cmu.edu}

\author[0000-0003-1227-3738]{Josiah Purdum}
\email{jpurdum@caltech.edu}
\affiliation{Caltech Optical Observatories, California Institute of Technology, Pasadena, CA 91125, USA}

\author[0000-0002-9267-6213]{Jillian C.~Rastinejad}
\email{jcrastin@umd.edu}
\altaffiliation{NHFP Einstein Fellow}
\affiliation{Department of Astronomy, University of Maryland, College Park, MD 20742, USA}

\author[0000-0002-0387-370X]{Reed Riddle}
\email{riddle@caltech.edu}
\affiliation{Caltech Optical Observatories, California Institute of Technology, Pasadena, CA 91125, USA}

\author[0000-0001-7648-4142]{Ben Rusholme}\email{rusholme@ipac.caltech.edu}
\affiliation{IPAC, California Institute of Technology, 1200 E. California Blvd, Pasadena, CA 91125, USA}

\author[0000-0003-1546-6615]{Jesper Sollerman}\email{jesper@astro.su.se}
\affiliation{The Oskar Klein Centre, Department of Astronomy, Stockholm University, AlbaNova, SE-10691 Stockholm, Sweden}

\author[0000-0001-8426-5732]{Jean J. Somalwar}
\email{jsomalwar@berkeley.edu}
\affiliation{Department of Astronomy, University of California, Berkeley, CA 94720-3411, USA}
\affiliation{Kavli Institute for Particle Astrophysics and Cosmology, Stanford, CA 94305, USA}

\author[0000-0002-3158-6820]{Sylvain Veilleux}
\email{veilleux@umd.edu}
\affiliation{Department of Astronomy, University of Maryland, College Park, MD 20742, USA}
\affiliation{Joint Space-Science Institute, University of Maryland, College Park, MD 20742, USA} 

\begin{abstract}

Tidal disruption events (TDEs) have traditionally been discovered in optical sky surveys through targeted searches of nuclear transients. However, it is expected that some TDEs will occur outside the galaxy nucleus, arising from wandering black holes originating in galaxy mergers. Here we present observations of \name, the first optical TDE discovered in the outskirts of a host galaxy. The TDE was identified by a custom `off-nuclear' implementation of the ML classifier \tdes, which classifies new ZTF transients based on their lightcurves. Follow-up observations confirm that \name is a TDE-H+He, occurring 9.5$"$ (9.3 kpc projected distance) from the nucleus of a massive galaxy (\Mstar= \galaxymass) with a central black hole mass of \bhmass. \name itself was likely disrupted by a much lighter black hole (\tdemass, as estimated with peak luminosity scaling relations). The black hole was either dynamically ejected from the nucleus or lies at the center of a very faint tidally-stripped dwarf galaxy undergoing a minor merger. Late-time observations of \name could confirm the origin of this apparent `wandering' black hole. The rate of highly offset ($\gtrsim$3 kpc) TDEs can be constrained to $<$10\% of the nuclear TDE rate, but our discovery implies that many dozens of similar sources will be detected by the Vera C. Rubin Observatory each year with resolvable offsets. 
\end{abstract}



\section{Introduction}

Tidal disruption events (TDEs) occur when stars pass close to massive black holes (BHs) \citep{rees_88}, generating a luminous flare that is detectable across the electromagnetic spectrum \citep[see][for a review]{gezari_21}. They offer a unique probe of otherwise-quiescent BHs, providing a window into BH demographics, and accretion physics.  While the first candidate TDEs were identified via X-ray emission, the bulk of the $\sim$200 known TDEs \citep[see e.g.][for a recent compilation]{otter_25} are now found via time-domain optical surveys such as the Zwicky Transient Facility \citep[ZTF;][]{ztf_survey,ztf_science,ztf_obs}. It is widely accepted that massive BHs lie at the heart of most if not all galaxies \citep{kormendy_13}, and searches for TDEs have therefore been naturally focussed on studying `nuclear' transients \citep[e.g.][]{van_velzen_11,hung_18,ztf_tde_1,final_season, yao_23, dgany_23,masterson_24,somalwar_25,grotova_25,zhang_25}. However, given that mergers are thought to contribute substantially to the growth of galaxies, a single galaxy could host many additional massive BHs. A smoking-gun signature for these additional BHs would therefore be the detection of a TDE with a measurable offset from the center of a galaxy \citep[see e.g.][]{ricarte_21b}. 

Several off-nuclear TDEs have now been found. The earliest candidate was \wings \citep{maksym_13,maksym_14}, discovered via its X-ray emission, with a faint detected underlying host (M$_{V} = -14.8$) that is a likely dwarf galaxy. More recently, \xmm was identified via its X-ray emission \citep{3xmm} and was located in a faint extended source thought to be a globular cluster or ultracompact dwarf galaxy. Another X-ray selected TDE candidate is EP240222A /  
AT 2024agqv \citep{ep240222a}, identified by the Einstein Probe \citep[EP;][]{ep_22}. This TDE had a very faint transient optical counterpart (M$_{g}=-14.6$), and was also located in a tiny satellite galaxy offset 35 kpc from a large neighbouring galaxy at the same redshift \citep{ep240222a}. Amongst the recent sample of 31 X-ray selected TDEs from eROSITA \citep{grotova_25}, one source (\erass) was located in a merging system offset 18$"$ (21 kpc) and 24$"$ (28 kpc) from the centers of two galaxies with an accompanying faint optical flare. Another recent candidate is HLX-1 in NGC 6099, a X-ray selected candidate intermediate-mass black hole (IMBH)-TDE \citep{chang_25}. 

There has also been one confirmed optically-selected source: off-nuclear TDE 2024tvd \citep{yao_25}. This TDE was discovered by ZTF \citep{tvd_disc} and initially classified as a classical TDE owing to its roughly nuclear location \citep{tvd_class}. However, multi-wavelength observations at high spatial resolution confirm that the TDE itself is located 0.9$"$ (0.8 kpc) from the center of its host galaxy \citep{yao_25}. Replicating this discovery remains challenging, because similar offsets could only be resolved for TDEs occurring in nearby galaxies, and they require expensive follow-up for confirmation. A more recent candidate was AT 2024puz, an exceptional rapid transient of uncertain origin that was discovered in a search for constant-colour off-nuclear ZTF transients \citep{somalwar_25b}. The source exhibited properties lying between those of a luminous fast blue optical transient and a TDE, in contrast to TDE 2024tvd which clearly resembled a `classical' optical TDE in all respects.

The ZTF collaboration recently began a new search for off-nuclear TDEs by classifying all transients (nuclear, offset and hostless) using only lightcurve data, adapting the existing realtime scanning infrastructure built around the machine-learning classifier \tdes \citep{tdescore, tdescore_realtime}. We here present the first off-nuclear TDE identified by this search: \name. The source was identified as a candidate TDE by \tdes based on its lightcurve properties, and subsequent follow-up data confirms this classification. The source exhibits luminous UV and soft X-ray emission, as expected for a TDE, and spectroscopic observations also match known TDEs. 

The paper is organised as follows: Section \ref{sec:discovery} outlines the real-time search for off-nuclear TDEs, while Section \ref{sec:obs} outlines the multi-wavelength observations of \name. Section \ref{sec:analysis} outlines the data analysis and modeling of the transient and host, while Section \ref{sec:origin} considers the possible origin of the parent BH. Finally, we summarise our results and provide an outlook in Section \ref{sec:conclusion}. Throughout the paper, we assume a flat $\Lambda$CDM cosmology with h = 0.7, $\Omega_{M}$ = 0.3 and $\Omega_{\Lambda}$ = 0.7.

\section{Discovery with \tdes}
\label{sec:discovery}

\begin{figure}
    \centering
    \includegraphics[width=0.99\linewidth]{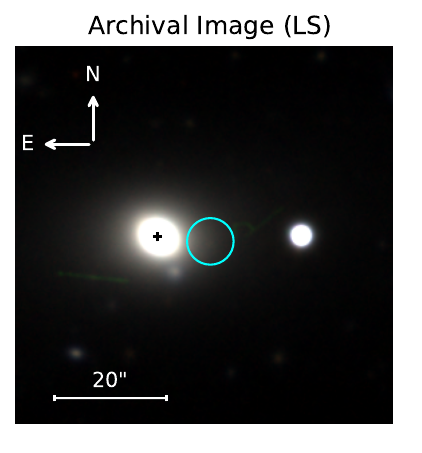}
    \includegraphics[width=0.99\linewidth]{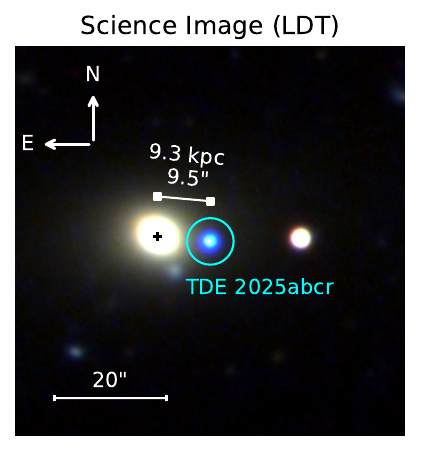}
    \caption{\textbf{Top:} Archival composite g/r/i image from Legacy Survey DR10, with the position of the transient (blue circle) and host galaxy (marked with the black + symbol). A higher-contrast image is shown in Figure \ref{fig:zoomedpicture}. \textbf{Bottom:} Composite u/g/i image of \name, taken with LDT on 2025 -12-14. The transient (circled in blue) is significantly offset (9.5$"$ / 9.3 kpc) from the center of the host galaxy \host.}
    \label{fig:prettypicture}
\end{figure}

\begin{figure}
    \centering
    \includegraphics[width=0.99\linewidth]{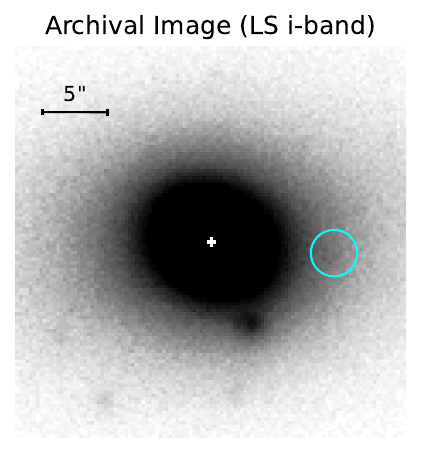}
    \caption{Zoomed-in archival i$-$band image from Legacy Survey DR10, centered on the host galaxy \host. The position of the transient (blue circle) and host galaxy (marked with the white + symbol) are shown. Four additional faint sources are visible south of the galaxy. However, no source is seen at the position of \name to a depth of $m =  23.78$ / $M = -12.78$ (see Section \ref{sec:scarlet} for more details), ruling out the possibility of a bright dwarf galaxy origin for \name.}
    \label{fig:zoomedpicture}
\end{figure}

While historical searches for optical TDEs have relied on simple algorithmic cuts \citep[see e.g.][]{ztf_tde_1,final_season,yao_23}, dedicated machine-learning classifiers have proliferated in recent years \citep{fleet_tde,tdescore,needle_24,alerce_tde_25,fink_tde_25,bhardwaj_25,zheng_25}. Within ZTF, the classifier \tdes \citep{tdescore} is now the primary means of identifying new TDEs. The classifier is a simple binary classifier, which combines high-level lightcurve properties (e.g. rise time, color at peak) with host galaxy properties (e.g. WISE $W1-W2$ color) and transient detection properties (e.g. average offset from host galaxy). Following the implementation of a real-time \tdes ranking framework in July 2024 \citep{tdescore_realtime}, the majority of new optical TDEs have been identified by \tdes. 

Since July 2025, the realtime \tdes framework has been expanded to encompass more agnostic searches for new TDEs. Two new versions of \tdes were developed: an unbiased classifier which uses lightcurve features and nuclearity but no specific information about the host itself, and a completely agnostic `off-nuclear' classifier that relies exclusively on transient lightcurve properties without any knowledge about a possible host. The latter classifier runs on every new ZTF transient, and is intended to capture `non-nuclear' TDEs which are either offset from a galaxy or have a host which is too faint to be recovered in the reference Pan-STARRS1 \citep[PS1;][]{panstarrs} catalogue. 

The implementation of a classifier agnostic to nuclearity presents additional classification and computational challenges. TDEs represent a tiny fraction ($\sim$2\%) of nuclear sources, being outnumbered by supernovae (SNe) and active galactic nucleus (AGN) flares. Even after removing all variables, TDEs represent only $\sim$8\% of all nuclear transients \citep{tdescore}. When broadening a search to encompass all transients, the contaminant rate increases even further. Most SNe have a significant offset from the host galaxy nucleus \citep[see e.g.][]{wang_97}, and within a flux-limited survey TDEs represent just $\sim$0.5\% of all transients \citep{ztf_bts_2}. 

Nonetheless, \tdes is able to identify TDEs with relatively high precision. 
From the implementation of real-time ranking on 2025-08-01 to 2026-03-01, there have been a total of 10 nuclear TDEs reported to TNS. Of these, one source (TDE 2025zmb, \citealt{2026chm_tns}) was not selected because it was at low galactic latitude (the ZTF TDE search is restricted to $|b| > 10$). Another source (TDE 2025alrm, \citealt{2025alrm_tns}) was missed because it is a repeating TDE peaking below 20th mag, and was only discovered by the deeper WFST survey. Of the remaining 8 sources (TDE 2025vjw, \citealt{2025vjw_tns}; TDE 2025aapf, \citealt{2025aapf_tns}; TDE 2025aarm, \citealt{2025aarm_tns}; TDE 2025abqg, \citealt{2025abqg_tns}; TDE 2025afvr, \citealt{2025afvr_tns}; TDE 2025ahbd, \citealt{2025ahbd_tns}; TDE 2025aljd, \citealt{2025aljd_tns}; TDE 2026amv, \citealt{2026amv_tns}; TDE 2026chm, \citealt{2026chm_tns}; and TDE 2026dmt, \citealt{2026chm_tns}), all were recovered by the off-nuclear \tdes classifier with a high score 

\name was first selected as a candidate TDE by the `off-nuclear' \tdes on 2025-11-04, and assigned for spectroscopic follow-up. In this case, the ML classification by \tdes was driven primarily by the high lower bound on blackbody temperature (T $> 10^{4.02}$ K), the poor fit to a Type Ia SN with \sncosmo \citep{sncosmo} ($\chi^{2}$ per d.o.f. = 4.5), and the best-fit temperature increasing with time rather than cooling (+87 K per day). A waterfall plot explaining the reasoning behind the \tdes classification is shown in Appendix Figure \ref{fig:shap}, while the full \tdes source page for \name is shown in Appendix Figure \ref{fig:scanpage}. 

Subsequent multi-wavelength follow-up observations confirmed that the source was indeed a TDE, with the source exhibiting the unique spectroscopic signatures, UV emission and transient soft X-ray emission which distinguish TDEs from SNe and other contaminants (see Section \ref{sec:classification} for more details). The classification was promptly reported to the community via a TNS classification \citep{tns_class} and astronote \citep{astronote_class}.

\name was notable because it was significantly offset (9.5"/9.3 kpc projected distance) from the center of its apparent host (see Figures \ref{fig:prettypicture} and \ref{fig:zoomedpicture}), located in the outskirts of a massive galaxy of known redshift. For this reason, \name was not identified by any other ongoing TDE search effort. This is the first TDE found by the `off-nuclear' \tdes that was not also identified by the `classic' or `unbiased' selections. 

\section{Observations of \name}
\label{sec:obs}
We here outline the various observations of \name. Photometric observations are summarised in Figure \ref{fig:lc}, and listed in Appendix Table \ref{tab:photometry}. Spectroscopic observations are presented in Figure \ref{fig:spec} and listed in Appendix Table \ref{tab:spec}. The follow-up was coordinated and data uploaded through the \url{fritz.science} instance of \texttt{Skyportal} \citep{skyportal,skyportal_2}.

\begin{figure*}
    \centering
    \includegraphics[width=0.95\linewidth]{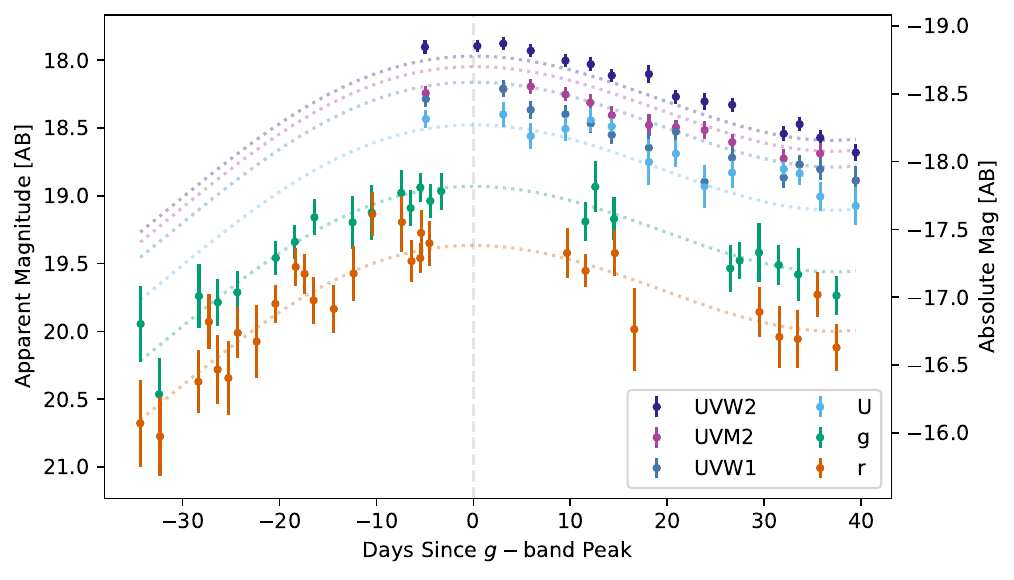}
    \includegraphics[width=0.48\linewidth]{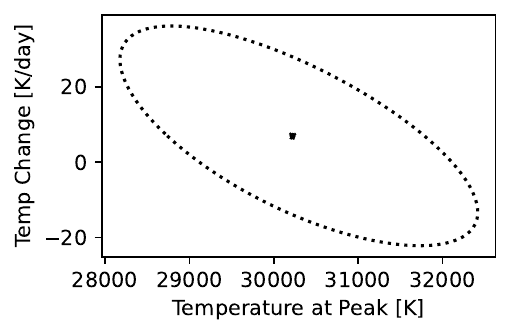}
    \includegraphics[width=0.48\linewidth]{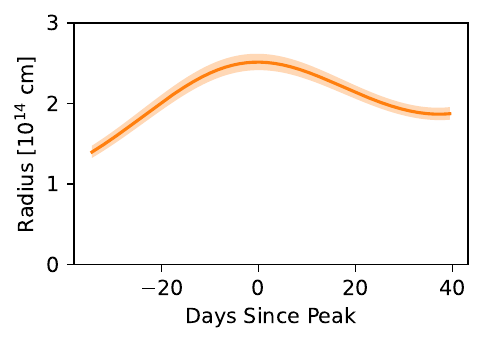}
    \caption{\textbf{Top:} Extinction-corrected UV/optical/NIR lightcurve of \name, alongside a blackbody fit to the data described in Section \ref{sec:lc}. The lightcurve is well described by a single black body with a linearly-evolving temperature. \textbf{Lower Left:} 1$\sigma$ uncertainty contour for the best-fit peak temperature and cooling rate. The temperature appears to change very little over time. \textbf{Lower Right:} Inferred blackbody radius as a function of time. The shaded region corresponds to the envelope of inferred radii when sampling the 1$\sigma$ peak temperature/cooling ellipse.}
    \label{fig:lc}
\end{figure*}

\subsection{Optical Observations}

On 2025-10-13, ZTF first detected a new optical transient, assigned the internal name ZTF25abxvhmk, in the footprint of the Northern-sky public survey. The source was reported to TNS as a transient on 2025-10-21, and assigned the designation AT 2025abcr \citep{tns_disc}. Observations were processed by the standard ZTF data reduction pipeline \citep{ztf_data}, and additional detections were recovered with the ZTF alert forced photometry\footnote{\url{https://zwickytransientfacility.github.io/ztf-avro-alert/schema.html}}. The source was detected in both $g-$band and $r-$band with an exceptionally blue color ($g-r = -0.3$ mag). \name brightened over a period of four weeks to a $g-$band peak on \tpeak (see Figure \ref{fig:lc}), and has since been slowly fading.

The source was also imaged in the optical with the Large Monolithic Imager (LMI) of the 4.3m Lowell Discovery Telescope \citep[LDT;][PI: Stein]{ldt_12,ldt_14} on 2024-12-14. Images were reduced using the LMI pipeline built with \texttt{mirar} \citep{mirar}, using Gaia DR3 \citep{Gaia2021} for astrometric calibration. Photometric calibration was performed using PanSTARRS \citep[PS1;][]{panstarrs} for $g/r/i$ and SkyMapper \citep{skymapper_dr4} for $u-$band. A composite cutout of the $u/g/i$ images is shown in Figure \ref{fig:prettypicture}. 

\subsection{UV observations}
Observations of \name were requested with the \swift \citep{gehrels_04} for multiple epochs, beginning 5 days before peak. Imaging was conducted with the Ultra-Violet/Optical Telescope \citep[UVOT;][]{swift_uvot} in all UV filters ($u$/UVW1/UVM2/UVW2). The UVOT observations were downloaded and reduced using \uvotredux \footnote{\url{https://github.com/robertdstein/uvotredux}} \citep{uvotredux_v0.3.1}, a Dockerized open-source python package for automated UVOT data reduction using  \texttt{swifttools}\footnote{\url{https://github.com/Swift-SOT/swifttools}} and \texttt{HEASoft}\footnote{\url{https://heasarc.gsfc.nasa.gov/docs/software/lheasoft/}}. Photometry was extracted assuming a 5$"$ aperture centred on the position of the transient, with an offset 10$"$ aperture to estimate the background. These observations revealed a bright source at the position of \name, separated from the nearby host galaxy nucleus. 

\subsection{NIR Observations}
We observed \name in the near infrared (NIR) with the Three Channel Imager (3KK;\citealt{lang2016wendelstein}) instrument mounted on the 2.1 m Fraunhofer Telescope at Wendelstein Observatory \citep{2014SPIE.9145E..2DH} in the $J$ and $Ks$ bands. The CMOS data were reduced using a custom pipeline \citep{2002A&A...381.1095G, 2025arXiv250314588B}. Astrometric calibration of the images was performed using the Gaia EDR3 catalog \citep{Gaia2021, 2021A&A...649A...2L, gaiaEDR3}. Tools from the AstrOmatic software suite \citep{sextractor, scamp, 2002ASPC..281..228B} were used for the coaddition of each epoch's individual exposures. We calibrate against the 2MASS Catalog \citep{2mass}. We use templates from Visible and Infrared Survey Telescope for Astronomy \citep[VISTA;][]{2012A&A...548A.119C} and the Saccadic Fast Fourier Transform (\texttt{SFFT}; \citealt{hu_image_2022}) algorithm for image subtraction. The source was not significantly detected ($>$3 $\sigma$) in any of these observations, so we provide upper limits in Appendix Table \ref{tab:photometry}.

\subsection{High-Cadence Imaging}
We observed \name with the prototype \textit{Cerberus} high-speed imager mounted at the prime focus of the 200-inch Hale telescope at Palomar Observatory \citep{cerberus_mo_inprep}, to search for any evidence of rapid variability. Rapid minute-scale variability was unexpectedly observed for the luminous fast blue optical transient AT 2022tsd \citep{ho_23}, and can be used to constrain the spatial extent of the emitting region of a source. Our observations were motivated by the potential similarity of \name to IMBH-TDE EP240222A, given that a detection could provide a constraint on the underlying black hole mass. The observations were performed on 2025-11-24, lasting 10\,min in the SDSS-\textit{g} band at 1\,s cadence and 10\,min in the SDSS-\textit{u} band at 5\,s cadence. The data were reduced with a custom pipeline \citep{cerberus_mo_inprep}.
We detect \name clearly in individual exposures, with signal-to-noise (S/N) ratios of $\approx$ 30 in $g$ and $\approx 8$ in $u$. Using a Lomb-Scargle periodogram \citep{1976Ap&SS..39..447L, 1982ApJ...263..835S}, we find no significant evidence for periodic behaviour, with no peak exceeding the 1\% false alarm probability threshold. However, these observations do not preclude variability on other timescales, or occurring outside of our period of observations.

\subsection{Spectroscopic Observation}
\label{sec:spec}

\begin{figure*}
    \centering
    \includegraphics[width=0.95\linewidth]{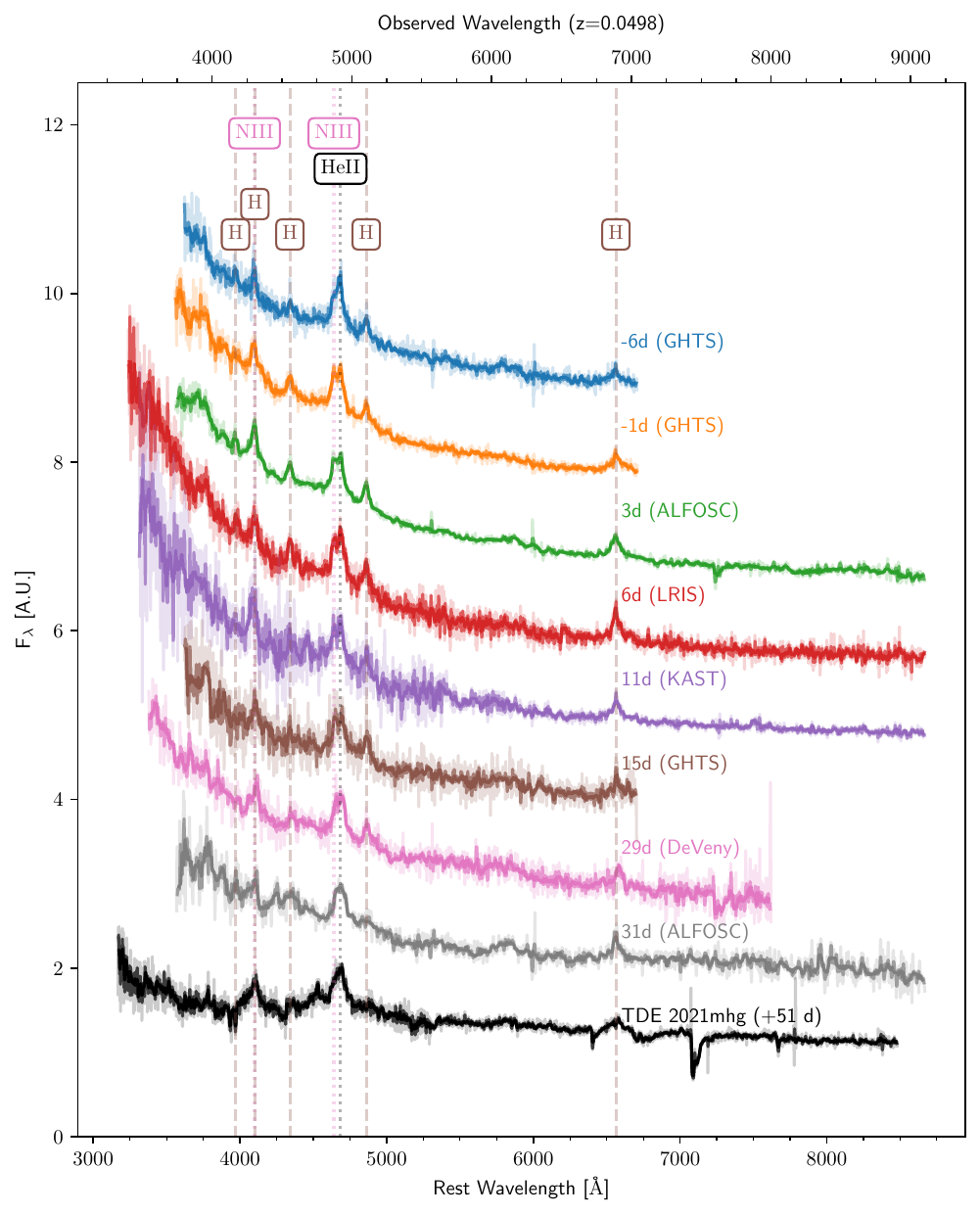}
    \caption{Spectral sequence of \name, with times given in rest frame days relative to the peak date. The source exhibits a blue continuum with little apparent cooling, broad H lines and a blended feature of He II ($\lambda4686$) and N III ($\lambda4640$). A comparison spectrum of TDE 2021mhg \citep{2021mhg_tns}, a TDE belonging to the same TDE-H+He spectral subclass at a similar phase, is shown in black at the bottom of the figure. The spectral features of \name resemble this and other known TDEs.}
    \label{fig:spec}
\end{figure*}

\subsubsection{Optical Spectroscopy of \name}
The first spectrum of \name was taken with the low-resolution P60/SED Machine \citep[SEDm;][]{sedm_1} on 2025-11-05 (PI: Stein), and reduced using the standard \texttt{pysedm} pipeline \citep{sedm_2, sedm_3}. The spectrum revealed a blue continuum with no obvious SN features, but the resolution was insufficient for a classification. Simultaneous photometry was derived through the automated pipeline, and is listed in Appendix Table \ref{tab:photometry}.

Additional spectra were obtained with the Goodman High Throughput Spectrograph \citep[GHTS;][]{Clemens2004} mounted on the 4.1 m Southern Astrophysical Research (SOAR) telescope on 2025-11-08 and 2025-11-30 (PI: Carney). 
These were reduced using the standard GHTS pipeline with \texttt{pypeit} \citep{pypeit:joss_arXiv, pypeit:zenodo}. 
An additional SOAR epoch was taken on 2025-11-14 (PI: Clemens) and reduced using the custom Python routine outlined in \citet{Kaiser+2021}.

A spectrum was taken using the Low-Resolution Imaging Spectrometer \citep[LRIS;][]{lris} on the 10 m Keck-I telescope on 2025-11-21 (PI: Chornock).
A spectrum was also taken using the Kast Double Spectrograph \citep[Kast;][]{kastref} on the 3~m Shane Telescope on 2025-11-26 (PI: Chornock).
Poor conditions during both the LRIS and Kast observations resulted in exposures with substantially varying signal-to-noise ratios that required flux normalization and weighting by the inverse variance before coaddition.
Both the LRIS and Kast data were reduced following standard procedures outlined by \cite{silverman2012}. Two spectra were taken with the Alhambra Faint Object Spectrograph and Camera (ALFOSC) mounted on the  the Nordic Optical Telescope (NOT) (PI: Charalampopoulos). The spectra were taken on 2025-11-17 and 2025-12-16, and were reduced using the PyNOT reduction pipeline\footnote{\url{https://github.com/jkrogager/PyNOT/}}. An additional spectrum was taken with the DeVeny spectrograph \citep{ldt_14} on LDT on 2025-12-14 (PI: Stein). The data were reduced using the DeVeney pipeline in \texttt{PypeIt} \citep{pypeit:joss_arXiv, pypeit:zenodo}. 

These spectra are shown in Figure \ref{fig:spec}. They confirm that the source had a blue continuum, with broad H and He emission lines at a consistent redshift of $z=0.05$. The spectral shape and features confirm the classification of \name as a tidal disruption event, and that it falls into the TDE-H+He subclass following the nomenclature introduced in \citet{ztf_tde_1}. The temporal evolution of key lines is shown in greater detail in Appendix Figure \ref{fig:speczoom}, and a summary of all observations is provided in Appendix Table \ref{tab:spec}.

\subsubsection{Host Galaxy Spectra}
A spectrum of the nucleus of the massive host galaxy, \host, was obtained using SOAR on 2025-12-08 (PI: Carney). The spectrum  was also reduced with the same \texttt{pypeit} routine as the transient spectra.
The spectrum is shown in Appendix Figure \ref{fig:hostspec}.
The host galaxy does not display any of the classic AGN signatures (Balmer lines,  [O III] at $\lambda4959$ / $\lambda5007$). However, there are several clear absorption features: Ca II at $\lambda3934$ / $\lambda3969$ and Na I ($\lambda5890$ / $\lambda5896$) consistent with the archival 6dF redshift of $z=0.049848 \pm 0.00015$ \citep{Heath2009}. 

\subsection{X-ray observations}
\label{sec:xrt}

\begin{figure}
    \centering
    \includegraphics[width=0.95\linewidth]{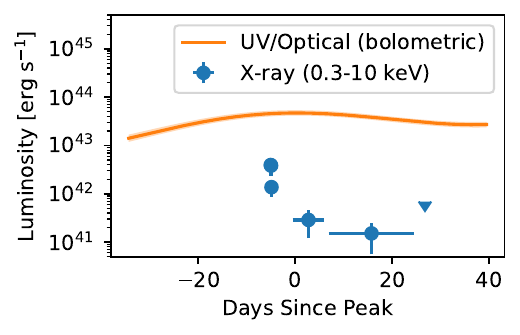}
    \caption{Bolometric luminosity (orange) from the UV/optical blackbody fit, alongside the X-ray detections (blue circles) and upper limits (blue triangles).}
    \label{fig:uvxray}
\end{figure}

We started observing \name with the X-Ray Telescope \citep[XRT;][]{swift_xrt} on board the \swift \citep{gehrels_04} on 2025-11-09T14:49:31 ($\delta t=$-4.82\,d,  PI: X. Hall). Here we present the analysis of the XRT observations acquired  until  2025-12-11T10:32:54 ($\delta t=25.48$\,d, PIs Stein, Guo; total exposure of 23.8\,ks). The observations are shown in Figure \ref{fig:uvxray}, and listed in Appendix Table \ref{tab:xray}.

We processed the  XRT data with HEASoft v6.36 and corresponding calibration files. An X-ray source is blindly detected at a position consistent with the optical transient in the first segment of observations at $\delta t=-4.82$\,d with count-rate $\approx 2\times 10^{-2}\,\rm{cts\,s^{-1}}$, significance of $4.5\sigma$ Gaussian equivalent.  The source is not blindly detected in individual observation segments at $\delta t>-4.6$\,d, implying very rapid fading of the X-ray emission and the physical association of the X-ray source with \name. 

Performing a targeted detection at the location of the X-ray transient following \citet{swift_online,Margutti13} leads to the recovery of a very faint X-ray source that persists until the end of our monitoring. By extracting the light-curves in two sub-energy bands (i.e., 0.3$-$1 keV vs. 1-10 keV), we find that the temporal evolution of the X-ray source is connected with the rapid fading of the very soft X-ray emission below the threshold of detectability for XRT, while the harder X-rays are consistent with a constant emission throughout the entire duration of our observations. With the angular resolution of the XRT we cannot rule out a dominant contribution of the host-galaxy emission to the harder X-rays. 

We extracted a spectrum using data taken between 2025-11-09T14:48:59.347 and 2025-11-09T23:01:48.401 (exposure of 2.9 ks). The spectrum is dominated by very soft X-ray emission. We fit the spectrum with an absorbed multitemperature disk spectrum (i.e., \texttt{tbabs*zashift*ezdiskbb} within \texttt{Xspec}). We employ W-statistics and use MCMC analysis to estimate the uncertainties of the inferred parameters.  We find no evidence for intrinsic neutral hydrogen absorption ($\rm{NH_{int}<0.5\times 10^{22}}\,\rm{cm^{-2}}$) and therefore fix the absorption to the Galactic value in the direction of \name ($\rm{NH}=1.55\times 10^{20}\rm{cm^{-2}}$; \citealt{HI4PI}). The metal abundances are set to Solar values with \texttt{wilm}. We find that the best-fitting model  under-estimates the emission at energies $>0.7$ keV. We thus update our model to include the contribution of a ``Comptonization tail'' (i.e., \texttt{tbabs*zashift*simpl*ezdiskbb}).\footnote{We note that adding the contribution of the Comptonization tail has minimal effects on the best-fitting temperature value and flux.} Within \texttt{simpl}  \citep{Steiner09} a hard power-law component with a spectral photon index $\Gamma$ is formed as a result of inverse Compton up-scattering of a fraction $f_{sc}$ of the initial thermal seed photon population. Due to the limited statistics, we adopt $\Gamma=2.5$ as found in other TDEs (e.g., \citealt{yao_25}), and fit the other parameters. We infer a disk maximum temperature $k_bT_{max}=41^{+5.0}_{-14}\,$eV and $f_{sc}=1.7^{+0.3}_{-1.5}\times 10^{-3}$. The 0.3$-$10 keV absorbed  (unabsorbed) flux is $3.5^{+0.7}_{-0.5}\times 10^{-13}\,\rm{erg\,s^{-1}cm^{-2}}$ ($4.6^{+0.9}_{-0.7}\times 10^{-13}\,\rm{erg\,s^{-1}cm^{-2}}$), corresponding to an unabsorbed luminosity of $L_{X} = 2.75^{+0.54}_{-0.42} \times 10^{42}$ erg s$^{-1}$.  

We conclude that \name initially shows the very soft X-ray emission that is a hallmark feature of TDEs, but this fades rapidly. Fast-fading X-ray emission was also observed for TDE 2019dsg \citep{bran} and TDE 2022dsb \citep{malyali_24}, and has been previously attributed to obscuration of the X-ray-emitting region by optically-thick stellar debris. However, the very rapid fading of the soft X-ray component of \name (a factor of $\sim$3 over the course of $\sim$4 hours) is exceptional. The X-ray variability of \name warrants further investigation that is beyond the scope of this paper, but a comparison to other known populations is explored in more depth in Section \ref{sec:classification}. 

\subsection{Radio Observations}

We observed the field of \name with the Karl G. Jansky Very Large Array (VLA) on November 12, 2025, roughly $2 \, {\rm days}$ before optical peak, under our dedicated program for off-nuclear TDEs (25B-109; PI: Sfaradi). This observation was conducted in X-band, with a central frequency of $10 \, {\rm GHz}$. We used 3C48 as an absolute flux and bandpass calibrator and J0204-1701 as a phase calibrator. We used the Common Astronomy Software Applications (CASA; \citealt{The_CASA_Team_2022}) packages and the VLA calibration pipeline to flag and calibrate the data, and applied additional flagging manually. We used the CASA task \texttt{TCLEAN} to produce clean images of the field, and the CASA task \texttt{IMSTAT} to calculate the image rms.

No radio emission is detected in the $10 \, {\rm GHz}$ band at the position of the TDE and at the center of the host galaxy down to a $3\sigma$ upper limit of $14 \, \mu {\rm Jy}$. This translates to a $3\sigma$ specific luminosity upper limit of $L_{\rm \nu} \leq 8.3 \times 10^{26} \, {\rm erg \, s^{-1} \, Hz^{-1}}$ at the distance of \name. For comparison, the first radio bright off-nuclear TDE\,2024tvd was initially not detected in the $10 \, {\rm GHz}$ band down to a similar $3\sigma$ depth of $L_{\rm \nu} \leq 8 \times 10^{26} \, {\rm erg \, s^{-1} \, Hz^{-1}}$, about $90 \, {\rm days}$ after optical discovery \citep{24tvd_Sfaradi_2025}. Late observations of TDE\,2024tvd revealed a late-time brightening and a peak luminosity of $L_{\rm \nu }\sim 10^{29} \, \rm erg \, s^{-1} \, Hz^{-1}$. Such re-brightening of the radio emission at late-times has been also been observed for numerous nuclear TDEs (e.g., ASASSN-15oi; \citealt{Horesh_2021}, AT\,2019azh; \citealt{Sfaradi_2022}, see also many examples in \citealt{Cendes_2024}). We therefore plan to continue monitoring \name with the VLA.

\section{Data Analysis}
\label{sec:analysis}

\subsection{Modelling of \name}
\subsubsection{UV/Optical Modelling}
\label{sec:lc}

We follow the lightcurve analysis procedure introduced in \citet{tdescore}, performing an iterative Gaussian process fit to the lightcurve data. We correct for extinction using results from \citet{schlafly_11} and the extinction law from \citet{fitzpatrick_99}. We model the multi-band lightcurve under the assumption that emission is thermal, and that it can be described by a blackbody spectrum with a temperature that evolves linearly in time \citep[see e.g.][]{ztf_tde_1}:

\begin{equation}
	T(t, k, T_{peak}) = k \times (t - t_{peak}) + T_{peak}
\end{equation}
where $t_{peak}$ is the time of peak, $T_{peak}$ is the temperature at peak and $k$ is the cooling rate. Using this temperature T, the flux can be derived for any observed frequency, $\nu$, and point in time, t:

\begin{equation}
	f(\nu, T_{peak}, k, t) = \frac{B(\nu, T)}{B(\nu_{0}, T)} \times GP_{\nu_{0}}(t)
	\label{eq:temp}
\end{equation}
where $B(\nu, T)$ is the blackbody spectral radiance,  $GP_{\nu_{0}}$ is the initial single-filter Gaussian Process fit and $\nu_{0}$ is the central frequency of the $g-$band filter. The best-fit values for $T_{peak}$ and k in Equation \ref{eq:temp} are found using \scipy minimisation \citep{scipy}. 

As seen in Figure \ref{fig:lc}, the data are well described by a blackbody  with a peak temperature of $T_{peak} = 30220 \pm 2120$ K and a varying inferred radius of order 10$^{14}$ cm. We infer a peak bolometric luminosity of $4.71^{+0.66}_{-0.59} \times 10^{43}$ erg s$^{-1}$ on \tpeak, which we adopt as the peak date. 

We can also estimate the mass of the BH which disrupted the star using the relation of \citet{mummery_24}:

\begin{equation}
    \log(\frac{M_{BH}}{\Msol}) = 6.52 + 0.98 \log(\frac{L_{\textup{g,peak}}} {10^{43}\textup{erg s}^{-1}})
\end{equation}

with an intrinsic scatter of $\sim0.53$ dex. Given the observed peak g$-$band luminosity of $\nu F_{\nu} = (3.66 \pm 0.34) \times 10^{42}$ erg s$^{-1}$, we estimate a BH mass of \tdemass.

\subsubsection{Comparison with known TDEs}
\begin{figure*}
    \centering
    \includegraphics[width=0.49\linewidth]{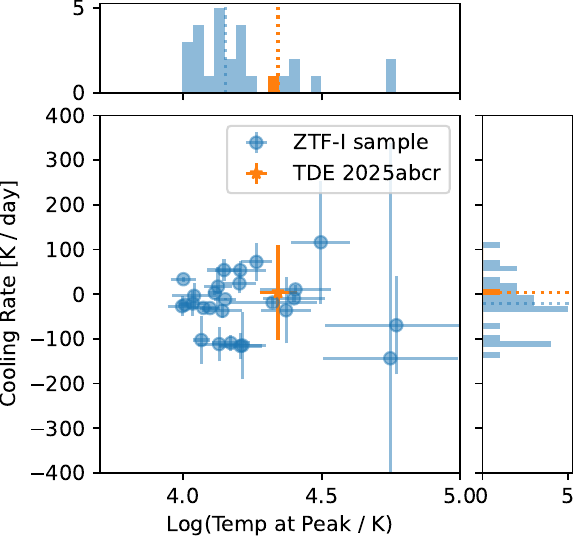}
    \includegraphics[width=0.49\linewidth]{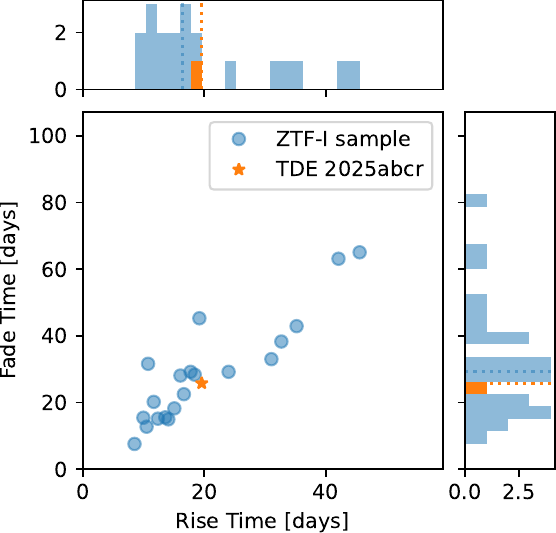}
    \includegraphics[width=0.49\linewidth]{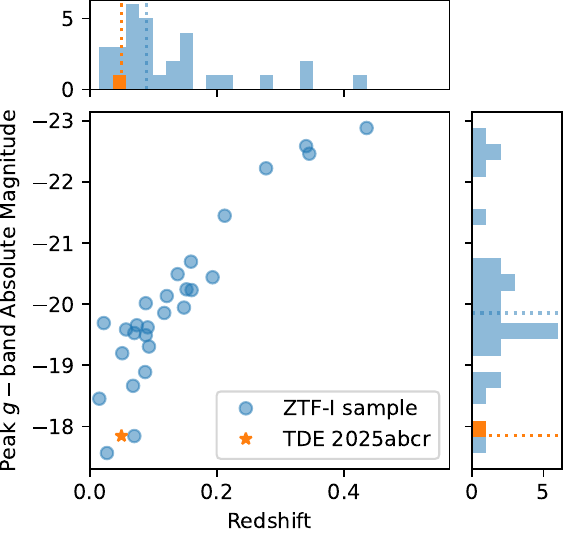}
    \includegraphics[width=0.49\linewidth]{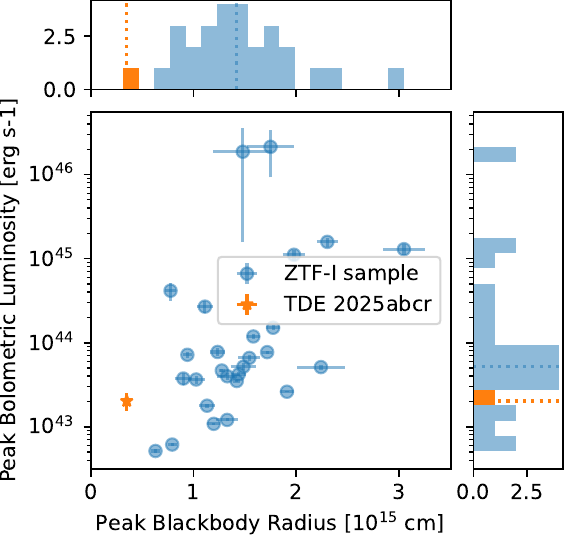}
    \caption{Comparison of the optical lightcurve properties of \name (in orange) and the sample of TDEs reported in \citet{final_season} (in blue). The median of each distribution is illustrated with a dashed line. The best-fit values for \name differ slightly from those in Figure \ref{fig:lc}, because no UV data was included in these fits. \textbf{Top Left:} Best-fit peak temperature versus cooling rate for TDEs, using the thermal model outlined in Section \ref{sec:lc} and fitting only ZTF photometric data. \textbf{Top Right:} Best-fit rise time and fade time for the same sample and fitting procedure, with the time defined as days between peak magnitude and 0.5 magnitude below peak. \textbf{Lower Left:} Redshift versus peak absolute $g-$band magnitude. \textbf{Lower Right:} Bolometric luminosity versus blackbody radius.  \name has a typical peak temperature and cooling rate, as well as a typical rise and fade time. In these lightcurve properties, the source is therefore an unremarkable optical TDE. However, the source has a low peak absolute $g-$band magnitude compared to other ZTF TDEs. As a relatively hot blackbody, \name has a much lower inferred blackbody radius than the overall ZTF population. This supports the interpretation that \name is a normal optical TDE but arises from a $\sim$$10^{6}$\Msol BH, as argued in more detail in Section \ref{sec:lc}.}
    \label{fig:sample_comp}
\end{figure*}

We can use the procedure outlined in Section \ref{sec:lc} to compare the properties of \name with other known TDEs. For consistency, we repeat the analysis of \name using only ZTF data, and then perform the same analysis for the sample of 30 TDEs discovered in ZTF-I \citep{final_season}. The best-fit values for \name therefore deviate slightly from those in Section \ref{sec:lc}, due to the absence of UV data. After performing the lightcurve fitting, we extract the key classifier features used by \tdes such as rise time and fade time. The results of these fits are shown in Figure \ref{fig:sample_comp}.

The general lightcurve properties of \name are consistent with the overall sample of nuclear TDEs in most key properties: inferred peak temperature and cooling rate, rise time and fade time. There is one clear difference: \name has a lower peak $g-$band absolute magnitude (M$_{g}$ = $-$17.6) than typical optical TDEs. There are other optical TDEs with comparable luminosities, but the source is clearly far below the population median. However, the peak magnitude remains significantly brighter than the transient optical counterpart (M$_{g}=-14.6$) to the X-ray-selected IMBH-TDE EP240222A/AT 2024agqv \citep{ep240222a}. \name thus qualitatively resembles a faint TDE from the broader optically-selected massive BH/TDE population, much more than it resembles EP240222A.

\subsection{Modelling of the host galaxy}
\subsubsection{Stellar Population Synthesis}
\label{sec:prospector}

We analyzed the host galaxy of \name, \host. We used \galsynthspec\footnote{\url{https://github.com/robertdstein/galsynthspec}} \citep{galsynthspec_v0.2.1}, an open-source Dockerized python package for automated galaxy SED fitting built using \prospector \citep{prospector} and \texttt{astroquery}. Photometry of the host galaxy was downloaded from GALEX \citep{galex}, SDSS \citep{sdss_00}, 2MASS \citep{2mass} and WISE \citep{wise}. The model provides a good fit to the galaxy SED. The resultant best-fit parameters are summarized in Appendix Table \ref{tab:prospector}, while the full corner plot is shown in Appendix Figure \ref{fig:prospector}. The most important conclusion from the fit is the tight constraint on the inferred surviving stellar mass of the galaxy: \Mstar = \galaxymass. From galaxy scaling relations of \citet{greene_20} for early-type galaxies, we would expect a central BH mass of \bhmass. The host galaxy is therefore very massive, much more so than typical TDE hosts \citep[see e.g.][]{final_season} but consistent with other off-nuclear TDEs (see Appendix Table \ref{tab:offnuclear_tde}). 

\subsubsection{Galaxy Profile Modelling}
\label{sec:scarlet}
In order to search for evidence of previous merger activity or the presence of an additional nuclear star cluster (NSC) offset from the host galaxy center, we undertook modeling of the coadded $g$, $r$ and $z$ band imaging available from the Legacy Survey Data Release 10 \citep{Dey2019} using the \texttt{Scarlet} multi-band scene modeling software\footnote{\url{https://pmelchior.github.io/scarlet/}} \citep{Melchior2018}. Note that the typical size of NSCs is 5\,pc \citep{Neumayer2020}, which corresponds to 6\,mas at $z=0.05$. If an off-nuclear NSC exists and is above the sensitivity limit, we only expect it to be detected as a point source. 

The LS DR10 has a $0.262^{\prime\prime}$ pixel scale and depths of $g\approx 24.7$, $r\approx 23.9$, $z\approx 23.0$\,mag. 
For the LS DR10 models, we provided \texttt{Scarlet} with the point spread function (PSF) model images provided by the data release. We ran \texttt{Source Extractor} \citep{sextractor} to identify all sources detected over a $5\sigma$ threshold. We required that all background galaxy models be monotonically decreasing --- but not radially symmetric --- and that they have the same morphology in each band (such that the SED does not vary in different regions of the galaxy). This enables us to avoid any assumptions about the galaxies following an analytical galaxy profile. For the primary host galaxy, we applied a double pseudo-S\'{e}rsic profile \citep{Spergel2010} as the stronger constraints on profile shape enabled improved modeling of the diffuse emission in the galaxy disk in comparison to the non-parametric model. \texttt{Scarlet} was run to convergence to fit the multi-band SEDs and galaxy morphologies for the sources in the scene. 

\begin{figure*}[htbp!]
\gridline{\fig{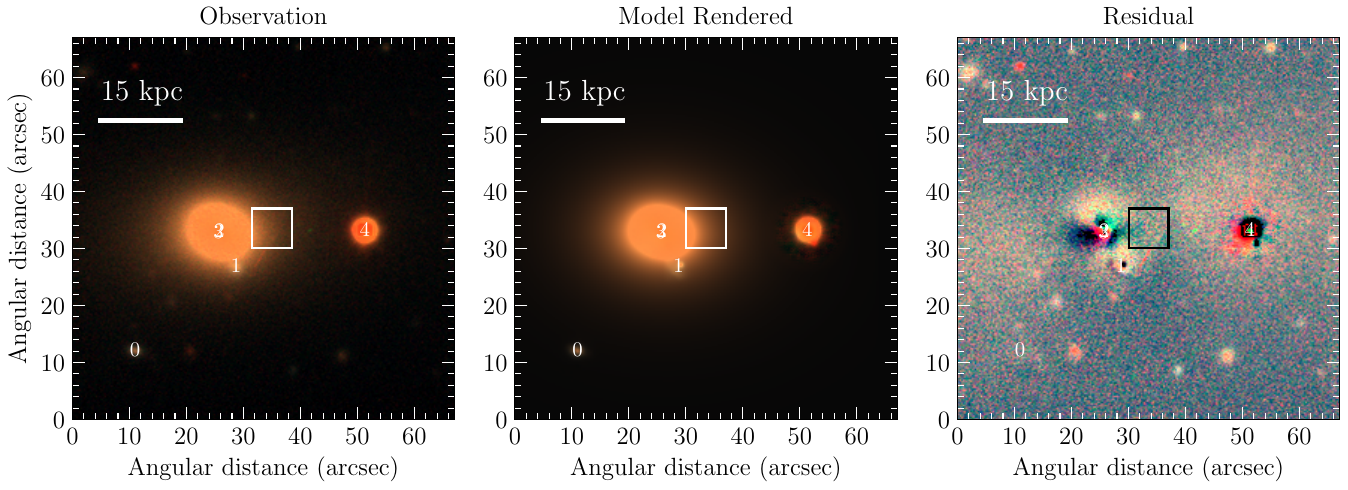}{0.99\textwidth}{}} 
\caption{\texttt{Scarlet} scene model from Legacy Survey DR10 imaging ($grz$). Each labeled source was modeled as a monotonically decreasing profile except for the host galaxy, which was modeled by a double pseudo-S\'{e}rsic in order to best account for the diffuse extended emission.  We show the model rendered to match the LS imaging (left), the coadded LS $grz$ image (center), and the data--model residual (right). The central box showed the region used to estimate an upper limit on a NSC cluster associated with the TDE position.}
\label{fig:Scarletmodel}
\end{figure*}

The best fit model, corresponding observations, and residuals are shown in Figure \ref{fig:Scarletmodel}. The LS residuals show evidence of a disturbed morphology in the galaxy bulge, but do not show any excess point-like emission at the location of the TDE. In the LS $g$-band image, we estimate the limiting magnitude  by determining the pixel variance of the residuals in a 30$\times$30 pixel cutout centered on the TDE position. We determine a 3$\sigma$ limiting magnitude of $g\approx 23.84$\,mag, which implies that no NSC exists at the TDE position with an absolute $g$-band magnitude brighter than $-12.78$\,mag.

Assuming a similar $g$-band mass-to-light ratio as EP240222A \citep{ep240222a}, which had a 10$^{7}$~\Msol\ dwarf host at an absolute magnitude of $-11.8$ mag, we find a limit of $<$10$^{7.4}~$\Msol\ for any dwarf galaxy at the position of \name. 

\subsubsection{BH Mass from velocity dispersion}

The Gaussian fit centroids of \name's broad H and He features closely agree with the redshift of the host \host, as derived from narrow galaxy lines. Using the broad H$\beta$ line, which is isolated and clearly resolved in all SOAR and LRIS spectra, we measure velocity differences between \name and \host to be negligible along our line of sight (mean 2 km s$^{-1}$, standard deviation 37 km s$^{-1}$). 

Using the spectrum of \host, we fit the stellar continuum with the \texttt{pPXF} package \citep{Cappellari2023} using the line masking recommendations outlined in \cite{Barth2002} and the \texttt{E-MILES} stellar population synthesis templates \citep{Vazdekis2016}. 
Prior to fitting we convolved the template spectra to match the instrumental resolution of GHTS (with a median $R\sim1430$ corresponding to a median $\sigma_{instrumental}\approx210$km $^{-1}$.
We measure a velocity dispersion of $\sigma = 324 \pm 14$ km $^{-1}$. Using the $M_\bullet$--$\sigma$ relation \citep{Ferrarese2000, Gebhardt2000} with updated coefficients from \cite{kormendy_13} we estimate the central BH mass of \host to be $\log(M_{BH}/M_\odot) = 9.41 \pm 0.31$, where the uncertainty includes the 0.29~dex intrinsic scatter from the $M_{BH}$--$\sigma$ relation. This is consistent with the independent estimate of \bhmass described in Section \ref{sec:prospector}.

\subsubsection{Is the host an active galaxy?}
We consider whether the host of \name is an active galaxy, because this would conclusively prove that at least one massive BH remains present in the galaxy nucleus. We see no evidence of this, after testing the following common AGN indicators:

\begin{itemize}
    \item There are no AGN-associated signatures in the host galaxy spectrum (e.g. [O III], broad Balmer lines).
    \item There is no detection of the galaxy in either our radio or X-ray observations, as would be expected for most AGN.
    \item There are no indications of previous nuclear variability in ZTF data.
    \item The WISE mid-infrared colors of the galaxy are not AGN-like ($W1 - W2 = -0.03$) \citep{stern_12}
\end{itemize}

While we cannot exclude the possibility of low-level AGN activity, we do not see any evidence supporting this. Instead, our observations suggest that \host is a quiescent early-type galaxy. Therefore, in contrast to the off-nuclear TDE 2024tvd where radio emission from the central BH was detected \citep{yao_25}, we cannot prove that a BH is still present at the center of the host galaxy of \name. 

\subsection{The nature of \name}
\label{sec:classification}
Given the observed properties of \name, we consider alternative origins for the transient. Beyond the interpretation of the source as an off-nuclear TDE, other potential explanations include an exotic supernova, a luminous fast blue optical transient (LFBOT) or an AGN flare.

Perhaps the most compelling evidence supporting a TDE nature is the detection of X-ray emission from the source (see Section \ref{sec:xrt}). Soft X-ray emission is a unique feature of TDEs, and can reliably distinguish them from AGN \citep[see e.g.][for a recent review of X-ray TDEs]{guolo_24}. The rapid X-ray fading over a timescale of hours is potentially reminiscent of a Quasi-Periodic Eruption (QPEs), a population that has now been associated with at least some TDEs \citep{nicholl_94}.
In contrast, while supernovae have been associated with luminous and rapidly-fading X-ray flashes \citep[see e.g.][for a summary of known events]{rastinejad_25}, these only occur within the first minutes after core collapse. Interacting supernovae can alternatively generate faint longer-lived X-ray emission over a period of weeks \citep[see e.g.][]{dwarkadas_14,chevalier_2017}. This X-ray emission is generally much less luminous than observed for \name (L$_X \sim 10^{40 - 42}$ erg s$^{-1}$), and moreover this emission has not been observed to rapidly vary on hour timescales \citep{chevalier_2017}. X-ray emission from interacting supernovae tends to be hard \citep[see e.g.][]{leising_94}, in stark contrast to the TDE-like soft emission we observe. LFBOTs can exhibit a range of X-ray properties \citep[see e.g.][]{margutti_19}, including very rapid variability, but the emission has never been observed which resembles the ultra-soft TDE emission with kT $\lesssim$1 keV. 

The photometric evolution of \name is also revealing. The source is extremely UV-luminous, with an inferred peak temperature of $\sim$30000K. The lightcurve evolution is essentially achromatic, with a negligible change in temperature observed over the course of the observations. This is a hallmark of TDEs \citep{ztf_tde_1}, and is indeed a key feature used to identify them in optical surveys \citep{tdescore}. While some interacting SNe can be UV-bright, the observed UV-optical colour of these transients is always observed to evolve over time \citep{jacobson_galan_24}. This behaviour is also inconsistent with expectations for SNe lacking circumstellar interaction, such as Type Ia, Type Ib/c and Type II SNe \citep[see e.g.][]{brown_09}. LFBOTs are characterised by luminous hot thermal emission similar to TDEs, but even the prototypical AT 2018cow exhibited significant cooling after peak \citep[see e.g.][]{perley_19}. 

The relatively low peak absolute magnitude of \name (M$_{g} =  -17.6$) is somewhat unusual for a TDE, with only 2/30 = 7\% of sources (TDE 2020ocn and TDE 2020wey) having a fainter peak in the comparison \citet{final_season} TDE sample. Interestingly, the luminosity lies between the two offnuclear TDEs for which an optical transient was detected (M$_{g} =  -19.6$ for TDE 2024tvd and M$_{g} =  -14.6$ for EP240222A). The peak would also be fairly underluminous for an interacting supernovae. Of the 123 Type IIn supernovae found by the ZTF Bright Transient Survey \citep{ztf_bts_1,ztf_bts_2}, only 16/123 = 13\% were fainter than our source.
\name is neither fast nor luminous, in contrast to expectations for an LFBOT. The faint peak is an order of magnitude fainter than any known LFBOT, and would be an extreme outlier for that population. The relatively slow evolution of \name (with a rise time of $\sim$30 days) would also be significantly (at least a factor of 3) slower than any other known LFBOT \citep[see e.g. Figure 4 of][]{lebaron_26}. 


We also consider the spectroscopic properties of \name. As detailed in Section \ref{sec:spec}, we observe broad H+He, measuring H$\alpha$ and He II FWHMs of $\sim$2500 km s$^{-1}$ for \name. TDEs exhibit considerable diversity in spectral properties, but approximately half of those discovered show similar H+He emission lines \citep{final_season}. However, the width of these lines tends to be broader than observed in our source, with a median H$\alpha$ FWHM of $\sim$10000 km s$^{-1}$ when fit with a single Lorentzian in a sample of 16 well-studied TDEs \citep{charalampopoulos_22}. Of this sample, 4/16=25\% of sources had comparable H$\alpha$ velocities and 1/16=6\% had comparable He II velocities at +30 days post-peak \citep{charalampopoulos_22}, highlighting that the relatively narrow features are not completely unprecedented. Off-nuclear TDE EP240222A also exhibited narrow lines with velocities of 1141 $\pm$ 32 km s$^{-1}$ and 1008 $\pm$ 32 km s$^{-1}$ for H$\alpha$ and He II respectively, a factor of $\sim2$ narrower than those measured for our source. Both \name and TDE EP240222A show broader H$\alpha$ emission lines than the narrow components of Type IIn SN studied in \cite{tadia_2013} and longer lived He II lines than seen in the flash features of young SN \citep{bruch_2021}. A comparison to these samples is also shown in Appendix Figures \ref{fig:h-alpha-in-context} and \ref{fig:he-II-in-context}.


To further quantify the resemblance of the spectroscopic features, we fit the LDT spectrum (+29 days post peak) using SNID \citep{snid}, for which the only match is a pre-peak spectrum of Type IIn supernova S1998s \citep{s1998s}. This spectrum was a classic example of a `flash ionisation' features \citep{shivvers_15}, in which early ionisation of the CSM interaction yields rapidly-fading narrow emission lines \citep{galyam_14}. However, these flash ionisation features do not typically last longer than $\sim$10 days \citep{khazov_16}, and to see them clearly a full $\sim$30 rest-frame days after peak would be completely unexpected. Our spectral observations are inconsistent with any alternative variety of supernova (Type Ia, Ib/c or II), and this is demonstrated clearly by the lack of any plausible additional matches identified by SNID.

One additional piece of evidence disfavours a supernova: the lack of any apparent star formation at the location of the transient. No source is detected at this position in archival GALEX AIS imaging in the FUV \citep{galex}. Assuming a point source exists at the detection threshold for this imaging ($\sim$20 mag AB), correcting for galactic extinction, and applying the UV-SFR scaling relations of \citet{hunter_10}, we place an upper limit of $<$0.3 \Msol year$^{-1}$.

To summarise, we exclude an interacting supernova origin for a variety of reasons:

\begin{itemize}
    \item The X-ray properties of \name (ultra-soft, transient, rapidly-varying, luminous and detected well after lightcurve onset) would be completely unprecedented for a supernova. 
    \item The optical spectra are only comparable to flash-ionised signatures observed in infant interacting supernova. However flash-features in supernovae fade within $\lesssim$10 days, while the lines are still present in our source at +30 rest-frame days from discovery. 
    \item The transient displays no apparent cooling over 70 days of data. While SNe can be UV-bright, the UVW2-g colour always evolves over these timescales.
\end{itemize}

We similarly exclude an LFBOT origin:

\begin{itemize}
    \item The peak absolute magnitude of \name is an order of magnitude fainter than any known LFBOTs
    \item The lightcurve for \name evolves far slower than any known LFBOT, with a rise/fade time of more than 60 days.
    \item The spectrum of \name (blue with broad H+He features) does not resemble any known LFBOT.
    \item The X-ray emission is ultra-soft, whereas LFBOT emission is harder.
\end{itemize}

We exclude an AGN origin:

\begin{itemize}
    \item No point source detected at the position of the transient, down to an absolute magnitude of M=-12.8, which would exclude any low-luminosity AGN.
    \item No evidence of archival radio emission at the location of the transient, or persistent X-ray emission.
    \item Transient, soft X-rays are inconsistent with an AGN origin
    \item Spectroscopic signatures associated with AGN (e.g. Balmer Lines, O[III]) are not present.  
\end{itemize}

In contrast, the TDE hypothesis explains the observations of \name well. The source is notably underluminous compared to the overall TDE population (see Figure \ref{fig:sample_comp}), but this is expected for an off-nuclear TDE given that wandering black holes should generally be less massive than central SMBHs. In all other respects (rise, fade, peak temperature and cooling),  the lightcurve of \name resembles a typical optical TDE. The only other unusual feature of this source is the relatively narrow H and He lines, which tend to be broader in other TDE-H+He. However, the source bears a strong spectroscopic resemblance to off-nuclear TDE EP240222A \citep{ep240222a}. To sum up, the evidence strongly disfavours any non-TDE origin for this transient but is consistent in almost all respects with expectations for a TDE. We therefore believe the classification of the transient as a TDE is robust.

\subsection{The rate of offset TDEs}
\name peaked in ZTF data with an apparent $g-$band magnitude of 18.9 mag, which was too faint to be selected by systematic transient classification programs such as the ZTF Bright Transient Survey \citep[BTS;][]{ztf_bts_1, ztf_bts_2}. However, given that the observed optical properties of \name (clear off-nuclear location, moderate luminosity, slow evolution) are similar to core-collapse SNe, it is very likely that brighter off-nuclear TDEs would have been classified by BTS. As no comparable transient was discovered before, we can conclude that \name-like transients must be rare.

As of December 2025, the ZTF BTS survey\footnote{\url{https://sites.astro.caltech.edu/ztf/bts/bts.php}} has identified 24 TDEs passing SN-like cuts amongst a larger sample of 4728 SN-like transients brighter than 18.5 mag, a regime in which spectroscopic completeness is $\sim$93\%. None of these 4728 transients resemble \name. We can conservatively place an upper limit at 90\% confidence on the expected number of highly offset TDEs in the BTS sample at N $<$ 2.3 assuming Poisson counting uncertainties \citep{gehrels_86}. We can therefore conservatively constrain the rate of high-offset TDEs such as \name to less than f $\lesssim$ 2.3/24 = 9.6\% of the `nuclear' TDE rate, where nuclear in this context means TDEs without a resolved offset in ZTF data. 

It is difficult to measure the overall rate of off-nuclear optical TDEs, because we expect that some will not have resolvable offsets in ZTF data. While ZTF can confidently resolve offsets $>$1", this would only be sufficient to resolve a 2024tvd-like distance of 0.8 kpc up to a redshift of z=0.04. Only 3/24 TDEs (12\%) lie within this redshift range, and for the remainder, the ZTF survey data alone would not be sufficient to measure an offset of 0.8 kpc. However, we can say that at least one of the 24 BTS TDEs was confirmed to be offset (TDE 2024tvd). We can therefore place a lower limit on the overall rate of offset TDEs. A single detection implies a lower limit at 90\% confidence of N $>$ 0.05. We can therefore say that at least f $\gtrsim$ 0.05/24 = 0.2\% of all TDEs are offset, though the true fraction may be substantially higher. Our limit is consistent with the fraction of off-nuclear X-ray-selected TDEs \citep{grotova_25}.

\section{The Origin of \name}
\label{sec:origin}
By definition, TDEs occur at the position of BHs. We can therefore be sure that a massive BH is present in the outskirts of the massive galaxy, possibly originating from a galaxy merger. Following a merger, dynamical friction should eventually lead to a compact supermassive BH binary \citep{begelman_80}, 
but this process can take billions of years. In the interim, tidal stripping will progressively consume the lighter galaxy, leaving an `orphan' BH with ever-fewer stars that remains bound to the larger galaxy and slowly spirals inwards. 

Ultimately the BHs will form a compact binary that can remain stable for long periods.  Any incoming massive BHs from new mergers that approach a pre-existing binary will undergo complex three-body interactions that lead to the ejection of the least massive BH. These binaries will draw closer and may merge via gravitational wave emission, detectable via pulsar timing arrays \citep[see][for a recent review]{taylor_25} and LISA \citep{lisa_17}. Even after an eventual binary merger, nuclear recoil following a merger could lead to the ejection of the new central BH \citep[see e.g.][]{Campanelli_07,blecha_16}.

Merging galaxies may host a pre-existing nuclear star cluster, while ejected BHs may also partially retain a nuclear star cluster \citep[see e.g.][]{komossa_08,stone_12,liu_13,khonji_25}. The intrinsic TDE rate will depend strongly on the presence of a star cluster accompanying the BH, and they may be considerably less frequent than for nuclear BHs \citep[see e.g.][]{komossa_08, stone_12,liu_13,li_2019}. Beyond these merger-related channels, globular clusters (GCs) formed in-situ may host IMBHs of order 10$^{3-4}$\Msol \citep{lutzgendorf_13} alongside numerous stars. It is therefore natural to expect that TDEs could occur in GCs \citep{ramirezruiz_09,chen_18,fragione_18b,tang_24}, and this would be an additional channel for off-nuclear TDEs.

To summarise, we consider the following formation channels for an offset BH:

\begin{itemize}
    \item A - an inspiralling BH from a major merger
    \item B - an inspiralling BH from a minor merger
    \item C - a BH from a previous merger that was dynamically ejected by three-body interactions in the galaxy nucleus
    \item D - a `recoiling' central BH ejected from a now-empty galaxy nucleus
    \item E - an IMBH hosted by a GC 
\end{itemize}

The key results from our modelling are that \name appears to be a TDE with a parent BH mass of \tdemass, with no evidence of an underlying dwarf galaxy or star cluster. The host galaxy is massive, with an inferred BH mass much higher than the TDE BH mass. Given this context, it is clear that both scenario A and D are not viable owing to significant mismatch between TDE-BH and host-BH mass. Moreover, the inferred central BH mass (\bhmass) from the host galaxy is so large that it is very unlikely to produce a TDE even with extreme spin values. 

For scenario E, GCs span a range of absolute magnitudes peaking at M$_{V}=-7.5$ and not typically exceeding M$_{V}= -11$ \citep{rejkuba_12}. We would therefore not expect to detect a GC in archival LS imaging, and our non-detection would therefore be consistent with the GC scenario. However, the inferred TDE-BH mass (\tdemass) is substantially larger than would be expected for a globular cluster. M-L scaling relations have been found to hold for GCs \citep{lutzgendorf_13}, and with our absolute magnitude limit of $-12.78$ we would expect a BH mass no more massive than $10^{4.8}$ \Msol. Using the upper end of observed globular clusters (M$_{V} \approx -11$), we would expect $\lesssim10^{4.3}$ \Msol. In both cases, there is considerable tension with the large inferred TDE-BH mass of \name, and we therefore consider scenario E to not be a viable explanation. 

For a recent minor merger, we can apply a similar argument, and conclude that our non-detection of a host would exclude a dwarf galaxy hosting a $10^{6}$ \Msol~BH. However, tidal stripping of stars following a merger will lead to a dimming of the galaxy over time, such that the central BH will be substantially heavier than expected based on stellar luminosity. 

We also highlight that, following a merger, galaxy morphology is expected to remain disturbed for $\sim$1 Gy \citep[see e.g.][]{conselice_14}. This is longer than the typical time for a compact SMBH binary to form, but much shorter than the time taken for the binary to ultimately merge \citep{begelman_80}. We would therefore expect disturbed morphology to remain visible for recent merger-related scenarios (Scenarios A, B or C), but for scenarios involving BH binary coalescence or direct BH formation (D and E) any disturbed morphology would be unrelated to the TDE-BH. That the morphology of the host galaxy is indeed visibly disturbed provides additional evidence supporting the viability of a merger or dynamical ejection origin (see Section \ref{sec:scarlet}).

In summary, a historical minor merger (Scenario B) remains a plausible origin, alongside Scenario C. It is also interesting to note that the host galaxy of the transient is massive, as this is consistent with predictions that the number of wandering BHs will scale linearly with galaxy halo mass \citep{ricarte_21a}. These arguments are summarised in Table \ref{tab:Scenarios}.

We can also consider potential offsets in redshift between the transient itself and the host itself, due to the motion of the transient. 
For the dynamical ejection scenario, a single BH can be ejected with velocities of $\sim1000$ km s$^{-1}$ \citep{3_body_slingshot}. 
A BH ejected from the nucleus would likely experience significant deceleration from dynamical friction before reaching a 9.3 kpc offset and/or could be travelling on a trajectory perpendicular to our line of sight. As such, the observed low velocity offset only provides as a weak constraint rather than evidence against a merger kick or dynamical three-body interaction. 

\begin{table*}[]
    \centering
    \begin{tabular}{c||c||c|c || c}
    &Scenario & Arguments In Favor & Arguments Against & Viable?\\
    \hline
    \hline
    A&Major Merger & Disturbed morphology of host & TDE-BH mass $<<$ Host-BH Mass & N\\
    && Natural reservoir of stars&&\\
    \hline
    B&Minor Merger & Explains TDE-BH Mass & No bright underlying host & Y \\
    && Disturbed morphology of host &&\\
    && Natural reservoir of stars&&\\
    && Consistent with other off-nuclear TDEs &&\\
    \hline
    C&Dynamical Ejecta MBH & Explains TDE-BH mass & No redshift offset & Y\\
    && Disturbed morphology of host & Unclear reservoir of stars & \\
    && Consistent with TDE 2024tvd &&\\
    \hline
    D&Recoiling MBH & No direct evidence of BH in nucleus & TDE-BH mass $<<$ Host-BH mass& N \\
    &&  & Reduced reservoir of stars & \\
    \hline
    E&Globular Cluster & Host should be too faint to detect & TDE-BH mass is too large & N \\
    && Natural reservoir of stars&& \\
    \end{tabular}
    \caption{Scenarios for the origin of the BH which generated \name. Given the large difference between the inferred TDE-BH mass (\tdemass) and the inferred galaxy BH mass (\bhmass), we can exclude a major merger and a recoiling MBH origin. The inferred TDE-BH mass is too large for a GC origin.  However, a minor merger and a dynamical ejection remain viable.}
    \label{tab:Scenarios}
\end{table*}

\section{Discussion and conclusion}
\label{sec:conclusion}

To summarise, \name is an ML-selected off-nuclear TDE with properties broadly consistent with typical nuclear optical TDEs. Our UV, X-ray and spectroscopic observations confirm unambiguously that the source is indeed a TDE rather than an AGN flare or SN. The parent BH must be massive (\tdemass), but there is no evidence of emission at this location in archival imaging. The BH may have therefore originated in a previous minor merger or a dynamical ejection following a nuclear three-body interaction.

Further late-time observations will shed more light on the properties of \name and its origin. In particular, the TDE-BH mass can be measured more precisely with observations of the late UV-optical plateau \citep{mummery_24} which appears ubiquitous in optical TDEs \citep{van_velzen_19}, and potentially through joint modelling of any late-time X-ray emission that might be detected \citep{guolo_25}. Moreover, the presence of any underlying host could be tested through deep late-time imaging at the position of \name, after the transient itself has faded. This would constrain the viability of the `minor merger'  origin channel for the BH. If no underlying source is found, we would instead favour a dynamical ejection origin for \name. 

With the discovery of \name, we can already consider patterns in the enlarged sample of off-nuclear TDEs discovered to date. A list of well-studied examples is presented in Appendix Table \ref{tab:offnuclear_tde}. All occur in very massive galaxies, which is consistent with theoretical predictions that the number of wandering BHs in a galaxy scale with the host halo mass \citep{ricarte_21a}. However, the other properties of the flares are grouped into distinct categories:

\begin{itemize}
    \item \textbf{Candidate IMBH-TDEs in dwarf galaxies:} so far these are mostly X-ray-selected TDEs which occur at the center of dwarf satellites. They exhibit very faint optical emission, consistent with expectations from scaling relations \citep{mummery_24} and simulations \citep{martire_25} that IMBH-TDEs should be intrinsically low-luminosity in the optical. This category includes EP240222a, \xmm, and HLX-1. It excludes \name owing to its bright optical emission and relatively high inferred TDE-BH mass. 
    \item \textbf{Low-Offset ($\lesssim$3 kpc) Optical TDEs:} TDE 2024tvd is the prototypical example of this group, and appears to be an unexceptional optical TDE with typical thermal emission. It is likely that, given the poor spatial resolving power of ground-based optical telescopes, a substantial fraction of these TDEs will not be identified as off-nuclear. Some of the known optical TDEs may in fact belong to this category.
    \item \textbf{High-Offset ($\gtrsim$3 kpc) Optical TDEs:} \name is the first example of this group. Much like low-offset optical TDEs, the multi-wavelength properties appear similar to optically-selected nuclear TDEs. We can state with confidence that high-offset optical TDEs must be intrinsically rare ($<10\%$ of the nuclear TDE rate).  
\end{itemize}

Additional low-offset optical TDEs could be identified with systematic multi-wavelength follow-up, though this will remain impractical to conduct for all newly identified TDEs. One simple alternative would be to target any TDEs occurring in a very massive galaxy with $M_{BH} \gtrsim 10^{8}$~\Msol, as has been proposed for searches targeting massive BH binaries \citep[see e.g.][]{mockler_23b}. More broadly, searches could focus on TDEs which have an inferred BH mass that deviates substantially from predictions of the TDE scaling relation of \citet{mummery_24}. Off-nuclear TDEs will originate from lighter BHs than their central host galaxy BH mass would imply, and therefore be underluminous. Both techniques would have reliably identified both TDE 2024tvd and \name as clear outliers. 

Though the astrometric precision of ground-based surveys such as ZTF are limited, averaging the position across the dozens or hundreds of individual detections can also improve performance. We find the offsets between the averaged spatial position of the $\sim$150 ZTF-detected TDEs and their PS1 hosts are well described by Gaussians of width just 0.11\arcsec\xspace in both RA and Dec (similar to estimates with early ZTF data by \citet{ned_stark}). Priority should therefore be given to sources with large offsets ($<$1\% of ZTF TDEs should have averaged offsets $>0.4$\arcsec). In particular, we highlight ZTF23aapyidj/TDE 2023mfm \citep{2023mfm_disc,2023mfm_class} as an underluminous TDE with a significant offset (0.6\arcsec) occurring in a very massive galaxy. TDE 2023mfm therefore appears to resemble TDE 2024tvd. This (and other known TDEs) may well be off-nuclear.

Finding additional high-offset TDEs like \name is much more straightforward. Photometric selection with \tdes is already effective in reducing the SN contamination, and the offset nature can be easily identified by the imaging in all-sky surveys. The very large projected offset of \name is encouraging for the off-nuclear TDE discovery prospects with the \textit{Vera C. Rubin Observatory} \citep{lsst}. $\mathcal{O}$(10 kpc) offsets should be easily resolvable ($>$1") even at a redshift of $z=1$. Given that Rubin is expected to identify thousands of nuclear TDEs each year \citep{lsst_tdes, karman_26}, we can expect many more off-nuclear TDEs will be discovered in the near future.

\begin{acknowledgments}
We thank Muryel Guolo, Dan Perley and Carl Rodriguez for fruitful discussion about off-nuclear TDEs.

Based on observations obtained with the Samuel Oschin Telescope 48-inch and the 60-inch Telescope at the Palomar Observatory as part of the Zwicky Transient Facility project. ZTF is supported by the National Science Foundation under Award \#2407588 and a partnership including Caltech, USA; Caltech/IPAC, USA; University of Maryland, USA; University of California, Berkeley, USA; Cornell University, USA; Drexel University, USA; University of North Carolina at Chapel Hill, USA; Institute of Science and Technology, Austria; National Central University, Taiwan, and the German Center for Astrophysics (DZA), Germany. Operations are conducted by Caltech's Optical Observatory (COO), Caltech/IPAC, and the University of Washington at Seattle, USA. 

SED Machine is based upon work supported by the National Science Foundation under Grant No. 1106171.

The Gordon and Betty Moore Foundation, through both the Data-Driven Investigator Program and a dedicated grant, provided critical funding for SkyPortal.

These results made use of the Lowell Discovery Telescope, owned and operated by Lowell Observatory

Some of the data presented herein were obtained at Keck Observatory, which is a private 501(c)3 non-profit organization operated as a scientific partnership among the California Institute of Technology, the University of California, and the National Aeronautics and Space Administration. The Observatory was made possible by the generous financial support of the W. M. Keck Foundation. 
The authors wish to recognize and acknowledge the very significant cultural role and reverence that the summit of Maunakea has always had within the Native Hawaiian community. We are most fortunate to have the opportunity to conduct observations from this mountain. 

A major upgrade of the Kast spectrograph on the Shane 3 m telescope at Lick Observatory, led by Brad Holden, was made possible through gifts from the Heising-Simons Foundation, William and Marina Kast, and the University of California Observatories. Research at Lick Observatory is partially supported by a generous gift from Google.

This work is based (in part) on observations made with the Nordic Optical Telescope, owned in collaboration by the University of Turku and Aarhus University, and operated jointly by Aarhus University, the University of Turku and the University of Oslo, representing Denmark, Finland and Norway, the University of Iceland and Stockholm University at the Observatorio del Roque de los Muchachos, La Palma, Spain, of the Instituto de Astrofisica de Canarias under NOT programmes 72-504. The NOT data presented here were obtained with ALFOSC, which is provided by the Instituto de Astrofisica de Andalucia (IAA) under a joint agreement with the University of Copenhagen and NOT. 

This work made use of data supplied by the UK Swift Science Data Centre at the University of Leicester.

This paper contains data obtained at the Wendelstein Observatory of the Ludwig-Maximilians University Munich. We thank Christoph Ries for carrying out the observations.
Funded in part by the Deutsche Forschungsgemeinschaft (DFG, German Research Foundation) under Germany's Excellence Strategy – EXC-2094/2 – 390783311.

The national facility capability for SkyMapper has been funded through ARC LIEF grant LE130100104 from the Australian Research Council, awarded to the University of Sydney, the Australian National University, Swinburne University of Technology, the University of Queensland, the University of Western Australia, the University of Melbourne, Curtin University of Technology, Monash University and the Australian Astronomical Observatory. SkyMapper is owned and operated by The Australian National University's Research School of Astronomy and Astrophysics. The survey data were processed and provided by the SkyMapper Team at ANU. The SkyMapper node of the All-Sky Virtual Observatory (ASVO) is hosted at the National Computational Infrastructure (NCI). Development and support of the SkyMapper node of the ASVO has been funded in part by Astronomy Australia Limited (AAL) and the Australian Government through the Commonwealth's Education Investment Fund (EIF) and National Collaborative Research Infrastructure Strategy (NCRIS), particularly the National eResearch Collaboration Tools and Resources (NeCTAR) and the Australian National Data Service Projects (ANDS).

The National Radio Astronomy Observatory (NRAO) is a facility of the National Science Foundation operated under cooperative agreement by Associated Universities, Inc. We thank the NRAO for carrying out the Karl G. Jansky Very Large Array (VLA) observation.

\textit{Note Added} - Shortly before this work was accepted, we became aware of a later preprint by \citet{patra_26}. The work reaches many similar conclusions to our own, and presents additional JWST data of \name. We also thank the authors for highlighting a typo an earlier version of this manuscript, with the projected offset incorrectly given as 10.3 kpc rather than 9.3 kpc.  

\end{acknowledgments}

\clearpage

\begin{contribution}

RS developed the \tdes scanning infrastructure, coordinated multi-wavelength follow-up and led writing of this manuscript. 
JC obtained the classification spectrum of \name and led spectroscopic analysis.
CW contributed the galaxy profile modelling. 
RM led the Swift-XRT data analysis.
XJH, MB, DG, BO and AP provided Wendelstein imaging and analysis.
IS contributed radio observations and analysis.
RS, JC, RM, XJH, IS, IA, RC, SG, YY, AA, MJG, EH and JJS discovered \name as members of the ZTFBH working group.
RS, SG, SBC, JR and SV contributed LDT imaging and spectroscopy.
JC, IA, AA and BK contributed GHTS spectroscopy.
RC and EH contributed Kast and LRIS spectroscopy.
PC contributed NOT spectroscopy.
GM and IC contributed high-cadence imaging and analysis.
EB, MJG, SLG, MMK, JP, RR, BR and JS contributed to the development and operation of ZTF. 
All authors contributed to the development of this work.


\end{contribution}

%
\facilities{PO:1.2m (ZTF), Hale (protoCerberus), Keck:I (LRIS), LDT (DeVeney, LMI), NOT (ALFOSC), PO:1.5m (SEDm), SOAR (Goodman), Swift (XRT, UVOT), VLA, WO:2m (3KK)}

\software{
\texttt{astroquery} \citep{prospector},
\texttt{emcee} \citep{emcee},
\texttt{HEASoft},
\galsynthspec \citep{galsynthspec_v0.2.1},
\texttt{mirar} \citep{mirar},
\prospector \citep{prospector},
\texttt{SCAMP} \citep{scamp},
\texttt{scarlet} \citep{Melchior2018},
\texttt{Source Extractor} \citep{sextractor},
\texttt{swifttools},
\tdes \citep{tdescore},
\uvotredux \citep{uvotredux_v0.3.1}
          }


\clearpage
\appendix

\begin{longtable}{c|c|c | c |c|c | c}
Date & Phase & Instrument & Filter & Mag & Mag Err & Limiting Mag \\
& [days] & & [AB] & [AB] & [AB] \\
\hline
2025-10-11T08:01:30.000 & -32.7d & ZTF & r & 20.71 & 0.32 & 19.93\\ 
2025-10-11T09:03:15.998 & -32.7d & ZTF & g & 19.99 & 0.28 & 19.48\\ 
2025-10-13T06:42:19.996 & -30.9d & ZTF & g & 20.51 & 0.27 & 20.22\\ 
2025-10-13T09:04:32.998 & -30.8d & ZTF & r & 20.81 & 0.29 & 20.11\\ 
2025-10-17T07:51:44.001 & -27.0d & ZTF & r & 20.41 & 0.23 & 20.00\\ 
2025-10-17T09:20:47.996 & -26.9d & ZTF & g & 19.79 & 0.24 & 19.79\\ 
2025-10-18T10:04:42.004 & -26.0d & ZTF & r & 19.96 & 0.20 & 19.69\\ 
2025-10-19T07:29:10.000 & -25.1d & ZTF & r & 20.31 & 0.25 & 20.21\\ 
2025-10-19T07:57:41.999 & -25.1d & ZTF & g & 19.83 & 0.17 & 20.29\\ 
2025-10-20T10:04:44.000 & -24.1d & ZTF & r & 20.38 & 0.27 & 19.97\\ 
2025-10-21T08:15:31.000 & -23.2d & ZTF & g & 19.76 & 0.16 & 20.27\\ 
2025-10-21T09:16:45.998 & -23.1d & ZTF & r & 20.04 & 0.19 & 20.08\\ 
2025-10-23T08:03:28.999 & -21.3d & ZTF & r & 20.11 & 0.27 & 19.99\\ 
2025-10-25T06:32:10.997 & -19.4d & ZTF & r & 19.83 & 0.14 & 20.28\\ 
2025-10-25T07:37:55.001 & -19.4d & ZTF & g & 19.51 & 0.13 & 20.41\\ 
2025-10-27T06:43:57.999 & -17.5d & ZTF & g & 19.39 & 0.12 & 20.33\\ 
2025-10-27T08:39:23.003 & -17.5d & ZTF & r & 19.56 & 0.14 & 19.99\\ 
2025-10-28T06:48:13.000 & -16.6d & ZTF & r & 19.61 & 0.15 & 20.00\\ 
2025-10-29T05:07:29.001 & -15.7d & ZTF & r & 19.80 & 0.17 & 19.96\\ 
2025-10-29T07:02:03.002 & -15.6d & ZTF & g & 19.21 & 0.13 & 20.26\\ 
2025-10-31T07:01:41.998 & -13.7d & ZTF & r & 19.87 & 0.17 & 20.01\\ 
2025-11-02T05:32:58.004 & -11.9d & ZTF & g & 19.24 & 0.19 & 19.37\\ 
2025-11-02T08:07:42.997 & -11.8d & ZTF & r & 19.61 & 0.21 & 19.48\\ 
2025-11-04T04:57:02.998 & -10.0d & ZTF & g & 19.17 & 0.20 & 19.06\\ 
2025-11-04T06:30:18.003 & -9.9d & ZTF & r & 19.17 & 0.16 & 19.27\\ 
2025-11-05T04:34:00.434 & -9.0d & SEDM & r & / & / & 19.15\\ 
2025-11-07T06:01:46.998 & -7.1d & ZTF & g & 19.03 & 0.17 & 19.21\\ 
2025-11-07T07:31:03.003 & -7.0d & ZTF & r & 19.23 & 0.22 & 19.26\\ 
2025-11-08T05:14:23.004 & -6.2d & ZTF & g & 19.14 & 0.14 & 19.51\\ 
2025-11-08T07:22:57.003 & -6.1d & ZTF & r & 19.52 & 0.15 & 19.72\\ 
2025-11-09T05:05:42.081 & -5.2d & SEDM & r & 19.49 & 0.11 & 20.23\\ 
2025-11-09T05:07:39.999 & -5.2d & ZTF & g & 18.99 & 0.11 & 19.73\\ 
2025-11-09T05:59:40.154 & -5.2d & SEDM & i & / & / & 17.99\\ 
2025-11-09T07:04:56.001 & -5.1d & ZTF & r & 19.31 & 0.17 & 19.57\\ 
2025-11-09T16:36:04.308 & -4.8d & UVOT & UVW2 & 18.01 & 0.05 & 21.92\\ 
2025-11-09T18:25:27.604 & -4.7d & UVOT & UVM2 & 18.36 & 0.06 & 22.22\\ 
2025-11-09T18:54:39.851 & -4.7d & UVOT & UVW1 & 18.37 & 0.06 & 21.38\\ 
2025-11-09T18:56:44.076 & -4.7d & UVOT & U & 18.50 & 0.06 & 21.00\\ 
2025-11-10T03:38:13.998 & -4.3d & ZTF & r & 19.38 & 0.17 & 19.61\\ 
2025-11-10T05:47:26.998 & -4.2d & ZTF & g & 19.09 & 0.13 & 19.93\\ 
2025-11-11T08:46:55.998 & -3.2d & ZTF & g & 19.02 & 0.14 & 19.31\\ 
2025-11-11T20:56:34.368 & -2.7d & Wendelstein 3KK & Ks & / & / & 20.04\\ 
2025-11-11T21:35:15.072 & -2.6d & Wendelstein 3KK & J & / & / & 20.47\\ 
2025-11-15T02:12:11.703 & 0.4d & UVOT & UVW2 & 18.00 & 0.05 & 22.24\\ 
2025-11-17T18:18:58.672 & 2.9d & UVOT & UVW1 & 18.30 & 0.06 & 21.42\\ 
2025-11-17T18:20:07.196 & 2.9d & UVOT & U & 18.46 & 0.09 & 20.51\\ 
2025-11-17T18:23:36.308 & 2.9d & UVOT & UVW2 & 17.98 & 0.05 & 22.20\\ 
2025-11-17T18:28:41.054 & 2.9d & UVOT & UVM2 & 18.33 & 0.06 & 22.10\\ 
2025-11-20T14:17:11.864 & 5.6d & UVOT & UVW1 & 18.45 & 0.07 & 21.21\\ 
2025-11-20T14:19:13.981 & 5.6d & UVOT & U & 18.62 & 0.09 & 20.57\\ 
2025-11-20T14:23:01.595 & 5.6d & UVOT & UVW2 & 18.04 & 0.05 & 22.02\\ 
2025-11-20T14:29:42.328 & 5.6d & UVOT & UVM2 & 18.31 & 0.06 & 22.00\\ 
2025-11-24T04:14:34.376 & 9.0d & UVOT & UVW1 & 18.48 & 0.07 & 21.16\\ 
2025-11-24T04:16:15.518 & 9.0d & UVOT & U & 18.57 & 0.09 & 20.57\\ 
2025-11-24T04:20:23.483 & 9.0d & UVOT & UVW2 & 18.11 & 0.05 & 21.96\\ 
2025-11-24T04:27:06.460 & 9.1d & UVOT & UVM2 & 18.37 & 0.05 & 22.10\\ 
2025-11-24T08:18:45.003 & 9.2d & ZTF & r & 19.46 & 0.18 & 19.79\\ 
2025-11-26T05:33:06.998 & 11.0d & ZTF & g & 19.24 & 0.14 & 20.01\\ 
2025-11-26T06:03:19.999 & 11.0d & ZTF & r & 19.59 & 0.12 & 20.02\\ 
2025-11-26T17:37:43.214 & 11.5d & UVOT & UVM2 & 18.43 & 0.06 & 21.93\\ 
2025-11-26T18:50:26.884 & 11.5d & UVOT & UVW1 & 18.55 & 0.07 & 20.89\\ 
2025-11-26T18:53:10.043 & 11.5d & UVOT & U & 18.51 & 0.09 & 20.54\\ 
2025-11-26T18:56:03.651 & 11.5d & UVOT & UVW2 & 18.14 & 0.05 & 21.76\\ 
2025-11-27T06:03:52.001 & 12.0d & ZTF & g & 18.98 & 0.19 & 19.19\\ 
2025-11-28T22:53:45.341 & 13.6d & UVOT & UVW1 & 18.64 & 0.07 & 21.44\\ 
2025-11-28T22:54:57.649 & 13.6d & UVOT & U & 18.55 & 0.09 & 20.61\\ 
2025-11-28T22:59:42.662 & 13.6d & UVOT & UVW2 & 18.22 & 0.05 & 22.20\\ 
2025-11-28T23:01:41.106 & 13.6d & UVOT & UVM2 & 18.53 & 0.07 & 21.66\\ 
2025-11-29T04:31:39.999 & 13.8d & ZTF & g & 19.22 & 0.16 & 19.55\\ 
2025-11-29T06:34:14.998 & 13.9d & ZTF & r & 19.46 & 0.17 & 19.54\\ 
2025-12-01T06:38:13.998 & 15.8d & ZTF & r & 20.02 & 0.31 & 19.21\\ 
2025-12-02T02:05:42.184 & 16.6d & SEDM & i & / & / & 19.38\\ 
2025-12-02T18:45:44.993 & 17.2d & UVOT & UVW1 & 18.73 & 0.12 & 20.32\\ 
2025-12-02T18:47:39.910 & 17.2d & UVOT & U & 18.82 & 0.17 & 19.87\\ 
2025-12-02T18:51:29.880 & 17.2d & UVOT & UVW2 & 18.21 & 0.07 & 21.19\\ 
2025-12-02T18:57:56.170 & 17.2d & UVOT & UVM2 & 18.60 & 0.08 & 21.32\\ 
2025-12-05T13:27:51.135 & 19.9d & UVOT & UVW1 & 18.61 & 0.07 & 21.29\\ 
2025-12-05T13:30:04.874 & 19.9d & UVOT & U & 18.75 & 0.10 & 20.55\\ 
2025-12-05T13:35:36.055 & 19.9d & UVOT & UVW2 & 18.38 & 0.05 & 22.23\\ 
2025-12-05T13:44:52.699 & 19.9d & UVOT & UVM2 & 18.61 & 0.06 & 22.26\\ 
2025-12-08T12:25:39.647 & 22.7d & UVOT & UVW1 & 18.98 & 0.13 & 20.39\\ 
2025-12-08T12:28:36.459 & 22.7d & UVOT & U & 18.99 & 0.16 & 20.13\\ 
2025-12-08T12:34:31.294 & 22.7d & UVOT & UVW2 & 18.41 & 0.06 & 21.61\\ 
2025-12-08T12:44:51.692 & 22.7d & UVOT & UVM2 & 18.64 & 0.07 & 21.89\\ 
2025-12-11T04:14:36.004 & 25.2d & ZTF & g & 19.58 & 0.18 & 19.97\\ 
2025-12-11T08:39:24.592 & 25.4d & UVOT & UVW1 & 18.81 & 0.08 & 21.21\\ 
2025-12-11T08:41:37.924 & 25.4d & UVOT & U & 18.89 & 0.11 & 20.55\\ 
2025-12-11T08:47:09.915 & 25.4d & UVOT & UVW2 & 18.44 & 0.05 & 22.10\\ 
2025-12-11T08:56:31.579 & 25.4d & UVOT & UVM2 & 18.72 & 0.06 & 22.28\\ 
2025-12-12T03:23:56.003 & 26.2d & ZTF & g & 19.53 & 0.14 & 20.23\\ 
2025-12-14T03:23:40.001 & 28.1d & ZTF & g & 19.47 & 0.22 & 19.46\\ 
2025-12-14T04:29:12.998 & 28.1d & ZTF & r & 19.89 & 0.18 & 19.89\\ 
2025-12-16T04:05:51.003 & 30.0d & ZTF & g & 19.56 & 0.15 & 20.13\\ 
2025-12-16T05:37:21.999 & 30.1d & ZTF & r & 20.07 & 0.23 & 19.84\\ 
2025-12-16T16:16:45.089 & 30.5d & UVOT & UVM2 & 18.85 & 0.07 & 21.40\\ 
2025-12-16T16:21:47.198 & 30.5d & UVOT & UVW1 & 18.95 & 0.08 & 21.49\\ 
2025-12-16T16:25:03.308 & 30.5d & UVOT & U & 18.87 & 0.09 & 20.83\\ 
2025-12-16T16:27:33.770 & 30.5d & UVOT & UVW2 & 18.65 & 0.05 & 22.36\\ 
2025-12-18T03:07:40.996 & 31.9d & ZTF & r & 20.09 & 0.22 & 19.70\\ 
2025-12-18T04:03:27.000 & 31.9d & ZTF & g & 19.63 & 0.20 & 19.92\\ 
2025-12-18T07:35:41.441 & 32.0d & UVOT & UVM2 & 18.89 & 0.07 & 21.49\\ 
2025-12-18T07:41:49.428 & 32.0d & UVOT & UVW1 & 18.85 & 0.07 & 21.73\\ 
2025-12-18T07:45:40.330 & 32.0d & UVOT & U & 18.90 & 0.08 & 20.98\\ 
2025-12-18T07:51:33.941 & 32.1d & UVOT & UVW2 & 18.58 & 0.05 & 22.50\\ 
2025-12-20T04:08:13.001 & 33.8d & ZTF & r & 19.76 & 0.16 & 19.99\\ 
2025-12-20T10:33:34.136 & 34.1d & UVOT & UVM2 & 18.81 & 0.08 & 21.15\\ 
2025-12-20T10:40:48.033 & 34.1d & UVOT & UVW1 & 18.89 & 0.08 & 21.40\\ 
2025-12-20T10:45:09.833 & 34.1d & UVOT & U & 19.07 & 0.11 & 20.74\\ 
2025-12-20T10:52:21.518 & 34.1d & UVOT & UVW2 & 18.68 & 0.06 & 22.27\\ 
2025-12-22T02:56:06.996 & 35.7d & ZTF & g & 19.78 & 0.14 & 20.47\\ 
2025-12-22T03:03:17.001 & 35.7d & ZTF & r & 20.15 & 0.18 & 20.42\\ 
2025-12-24T02:38:45.191 & 37.6d & UVOT & UVW1 & 18.98 & 0.11 & 20.77\\ 
2025-12-24T02:42:04.545 & 37.6d & UVOT & U & 19.14 & 0.14 & 20.44\\ 
2025-12-24T02:49:50.256 & 37.6d & UVOT & UVW2 & 18.79 & 0.06 & 22.03\\ 
2025-12-24T02:59:44.831 & 37.6d & UVOT & UVM2 & 19.01 & 0.09 & 21.57\\ 

\caption{Photometry of \name. Phase is defined in rest-frame days relative to the lightcurve peak. }
\label{tab:photometry}
\end{longtable}

\begin{table*}[]
\centering
\begin{tabular}{c|c|c}
Date & Phase & Instrument \\
& [days] & \\
\hline
2025-11-05T04:34:49.000 & -9.0d & SEDm \\ 
2025-11-08T00:59:20.000 & -6.3d & GHTS \\ 
2025-11-14T00:00:00.000 & -0.6d & GHTS \\ 
2025-11-17T22:48:00.000 & 3.1d & ALFOSC \\ 
2025-11-21T07:26:23.000 & 6.3d & LRIS \\ 
2025-11-26T06:15:50.000 & 11.0d & KAST \\ 
2025-11-30T01:12:46.000 & 14.6d & GHTS \\ 
2025-12-15T04:08:24.000 & 29.0d & DeVeny \\ 
2025-12-16T21:21:36.000 & 30.7d & ALFOSC \\ 

\end{tabular}
\caption{Spectroscopic observations of \name. Phase is defined in rest-frame days relative to the lightcurve peak. }
\label{tab:spec}
\end{table*}

\begin{figure*}
    \includegraphics[width=0.95\linewidth]{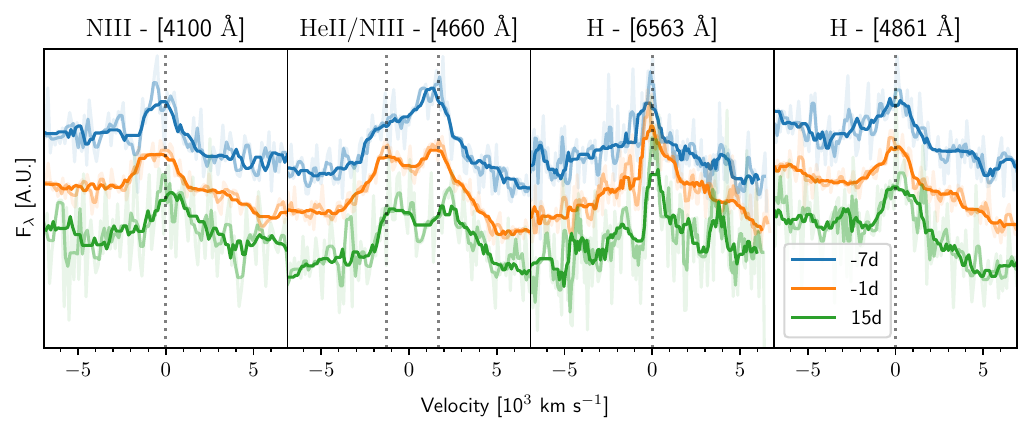}
    \caption{Zoomed-in velocity plots for an assortment of lines in the GHTS spectra of \name, with times given in observer frame relative to the peak date and the central wavelength given in square brackets. The blended He II ($\lambda4686$) / N III ($\lambda4640$) feature evolves over time, with the N III becoming more prominent. Each of these lines is marked by a vertical dashed line.}
    \label{fig:speczoom}
\end{figure*}

\begin{figure*}
    \centering
    \includegraphics[width=0.95\linewidth]{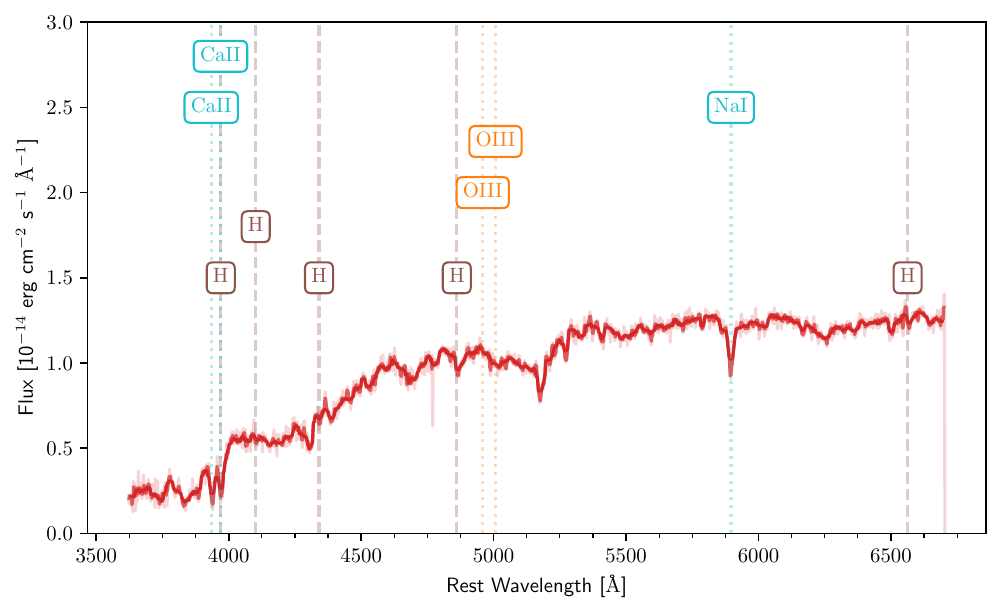}
    \caption{Spectrum of host \host, taken with GTHS. There is no evidence for typical AGN signatures (broad H or [O III]). However, clear absorption signatures are seen at Ca II  and Na I, for a consistent redshift of z = 0.0498.}
    \label{fig:hostspec}
\end{figure*}

\begin{table*}[]
\centering
\begin{tabular}{c|c|c|c|c}
UTC Start & UTC End & Phase & Window & Luminosity\\
& & [days] & [hr] & [$10^{41}$ erg s$^{-1}$] \\
\hline
2025-11-09T14:48:59.347 & 2025-11-09T14:57:20.807 & -4.82 & 0.14 & $39.4^{+17.7}_{-13.7}$\\ 
2025-11-09T16:31:25.310 & 2025-11-09T16:40:21.872 & -4.75 & 0.15 & $38.3^{+17.1}_{-13.2}$\\ 
2025-11-09T18:05:21.092 & 2025-11-09T23:01:17.524 & -4.60 & 4.93 & $13.7^{+5.6}_{-4.4}$\\ 
2025-11-14T05:13:41.900 & 2025-11-20T17:42:22.193 & 2.66 & 156.48 & $2.9^{+1.9}_{-1.4}$\\ 
2025-11-23T14:56:16.906 & 2025-12-11T07:26:20.761 & 16.94 & 424.50 & $1.5^{+1.1}_{-0.8}$\\ 
2025-12-11T10:04:27.529 & 2025-12-11T10:32:22.405 & 25.48 & 0.47 & $<$6.9\\ 

\end{tabular}
\caption{Dynamically-binned X-ray observations of \name. Phase is defined in rest-frame days from the bin midpoint to the inferred peak time. We provide the unabsorbed 0.3-10 keV luminosity, as outlined in Section \ref{sec:xrt}.}
\label{tab:xray}
\end{table*}

\begin{figure*}
    \centering
    \includegraphics[width=0.8\linewidth]{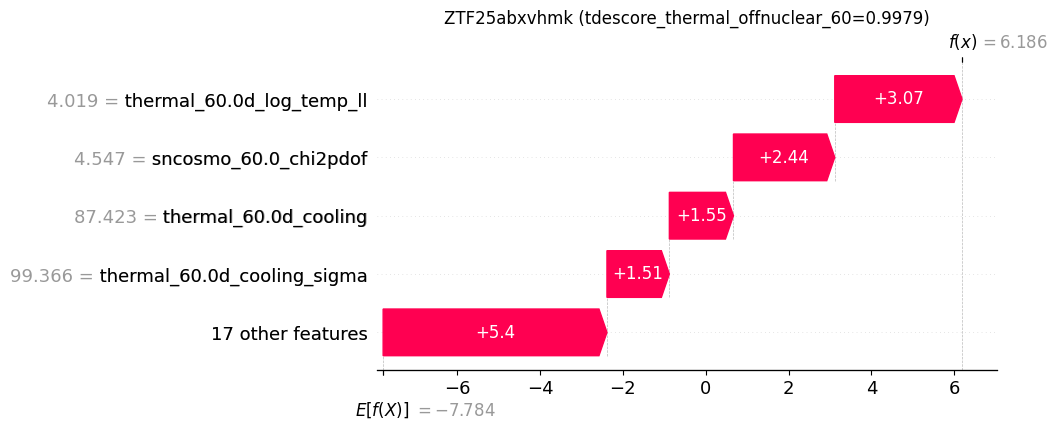}
    \caption{A waterfall plot highlights the main features which contribute to the ML classification of \name as a likely TDE by \tdes, as described in \citet{tdescore}. As of 2025 November 29, the classification is driven primarily by the high lower bound on blackbody temperature (T$ > 10^{4.02}$K), the poor fit to a Type Ia SN with \sncosmo \citep{sncosmo} ($\chi^{2}$ per d.o.f. = 4.5), and the best-fit temperature increasing with time rather than cooling (+87 K per day). All properties are characteristic of TDEs (which are hot, slowly evolving and do not rapidly cool), but are not common for SNe (which have lower temperatures, more rapid evolution and generally cool more quickly). The exact parameters differ slightly from Section \ref{sec:lc}, because they are based on a fit to only ZTF data and include only photometry available up to 2025 November 29.}
    \label{fig:shap}
\end{figure*}

\begin{figure*}
    \centering
    \includegraphics[width=0.99\linewidth]{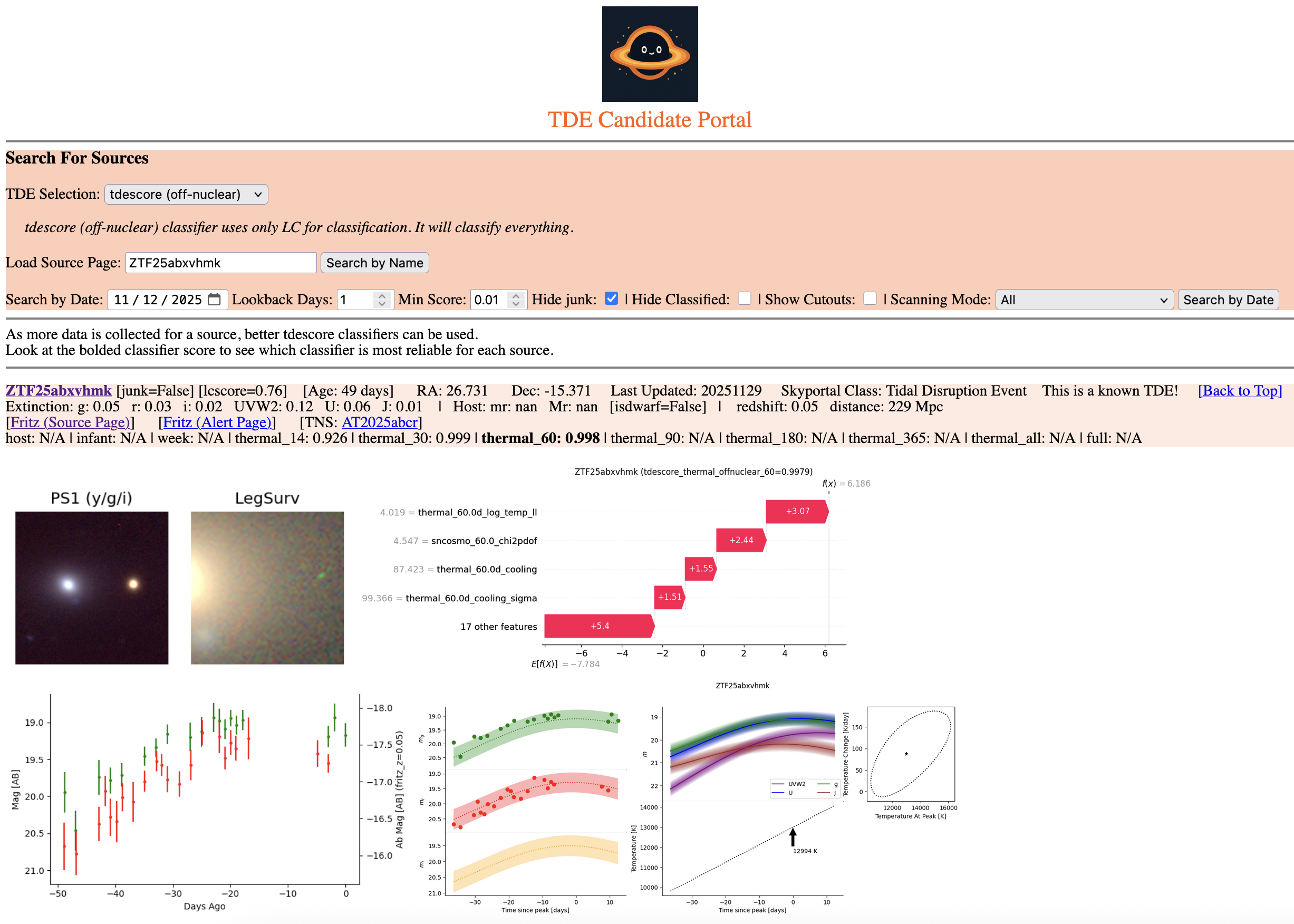}
    \caption{\tdes summary page for ZTF25abxvhmk / \name as of 2025-11-29. The classifier relies on the lightcurve fit shown in the central lower panel, following the same blackbody procedure described in Section \ref{sec:lc} and shown in the lower-right panel. The waterfall plot (center right) is reproduced in higher resolution in Figure \ref{fig:shap}.}
    \label{fig:scanpage}
\end{figure*}

\begin{table*}[]
\centering
\begin{tabular}{c|c|c|c|c|c}
Parameter & Best & Lower & Upper & Prior& Unit \\
		& Fit &Bound& Bound &  \\
\hline
tau & $1.05^{+0.10}_{-0.13}$ & 0.1 & 10.0 & Uniform  & Gyr$^{-1}$\\ 
tage & $8.98^{+0.80}_{-1.07}$ & 0.1 & 10.1 & Uniform  & Gyr\\ 
dust2 & $0.07^{+0.03}_{-0.03}$ & 0.0 & 1.0 & Uniform  & optical depth at 5500AA\\ 
logzsol & $0.19^{+0.01}_{-0.02}$ & -1.8 & 0.2 & Uniform  & $\log (Z/Z_\odot)$\\ 
log10(mass) & $11.39^{+0.02}_{-0.04}$ & 8.0 & 12.0 & LogUniform  & Solar masses formed\\ 
\hline 
log10(surviving mass) & $11.18^{+0.02}_{-0.03}$ & / & / & derived  & Solar masses surviving\\ 

\end{tabular}
\caption{Priors and best-fit values for the five parameters of the \prospector fit. The surviving solar mass parameter is also shown, but this is a derived quantity that does not directly enter the fit.}
\label{tab:prospector}
\end{table*}

\begin{figure*}
    \centering
    \includegraphics[width=0.9\linewidth]{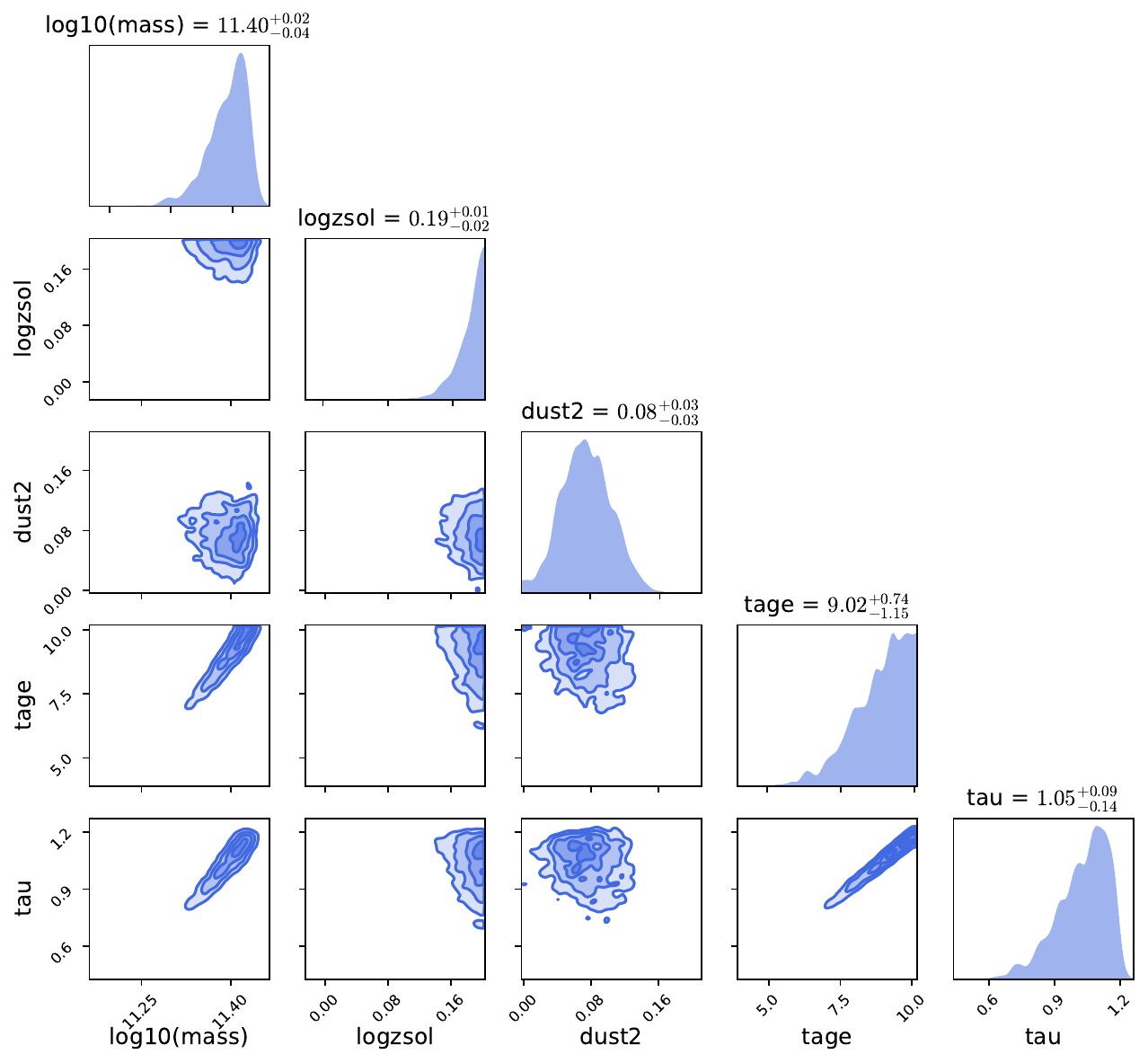}
    \includegraphics[width=0.65\linewidth]{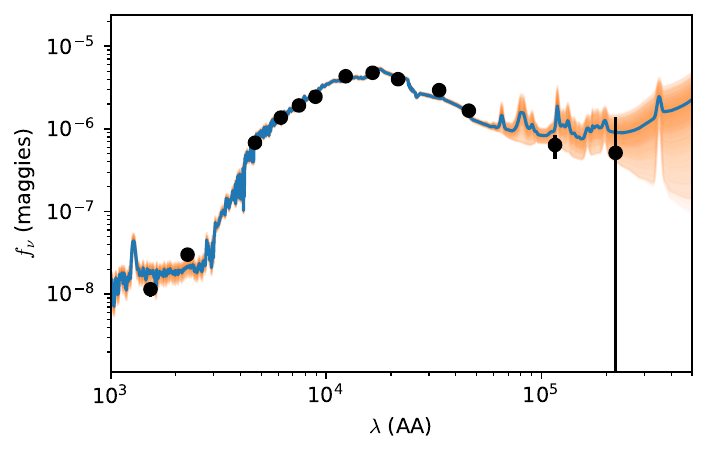}
    \caption{\textbf{Top:} Corner plot from the \prospector fit, with the priors listed in Table \ref{tab:prospector}. \textbf{Bottom:} Median SED (blue) and sampled uncertainty (orange) for the host galaxy, with measured photometry plotted in black. The \prospector model provides a good fit to the data across the full wavelength range, from the UV to mid infra-red.}
    \label{fig:prospector}
\end{figure*}

\begin{table*}[htbp!]
\centering
\label{tab:offnuclear_tde}
\begin{tabular}{ccccccc}
\hline
Name & $z$ & Offset & Parent $M_{\rm gal}$ & Satellite dwarf $M_{\ast}$ & Central $M_{\rm BH}$ & TDE $M_{\rm BH}$ \\
     &     & ($\arcsec$; kpc) & ($M_\odot$) & ($M_\odot$) & ($M_\odot$) & ($M_\odot$) \\
\hline
\xmm & 0.05526 & 11.6; 12.5 & $10^{10.93\pm0.07}$ & $10^{7.3\pm0.4}$ & $10^{8.49\pm0.67}$ & $\sim 10^{4.9}$ \\
EP240222a & 0.03275 & 53.1; 34.7 & $10^{10.89\pm0.07}$ & $10^{7.0\pm0.3}$ & $10^{8.44\pm0.67}$ & $\sim 10^{4.9}$ \\
TDE 2024tvd & 0.04494 & 0.92; 0.81 & $10^{10.93\pm0.02}$ & $<10^{7.6}$ & $10^{8.37\pm0.51}$ & $\sim 10^{5.9}$ \\
\hline
\name & 0.04985 & 9.5; 9.3 & $10^{11.18 \pm 0.03}$ & $<10^{7.4}$ & $10^{8.82 \pm 0.65}$ & $\sim 10^{6.1}$ \\
\hline
\end{tabular}
\caption{Summary of all off-nuclear TDEs, extended from the table of \citet{yao_25} with the inclusion of \name.}
\end{table*}

\begin{figure*}
    \centering
    \includegraphics[width=0.9\linewidth]{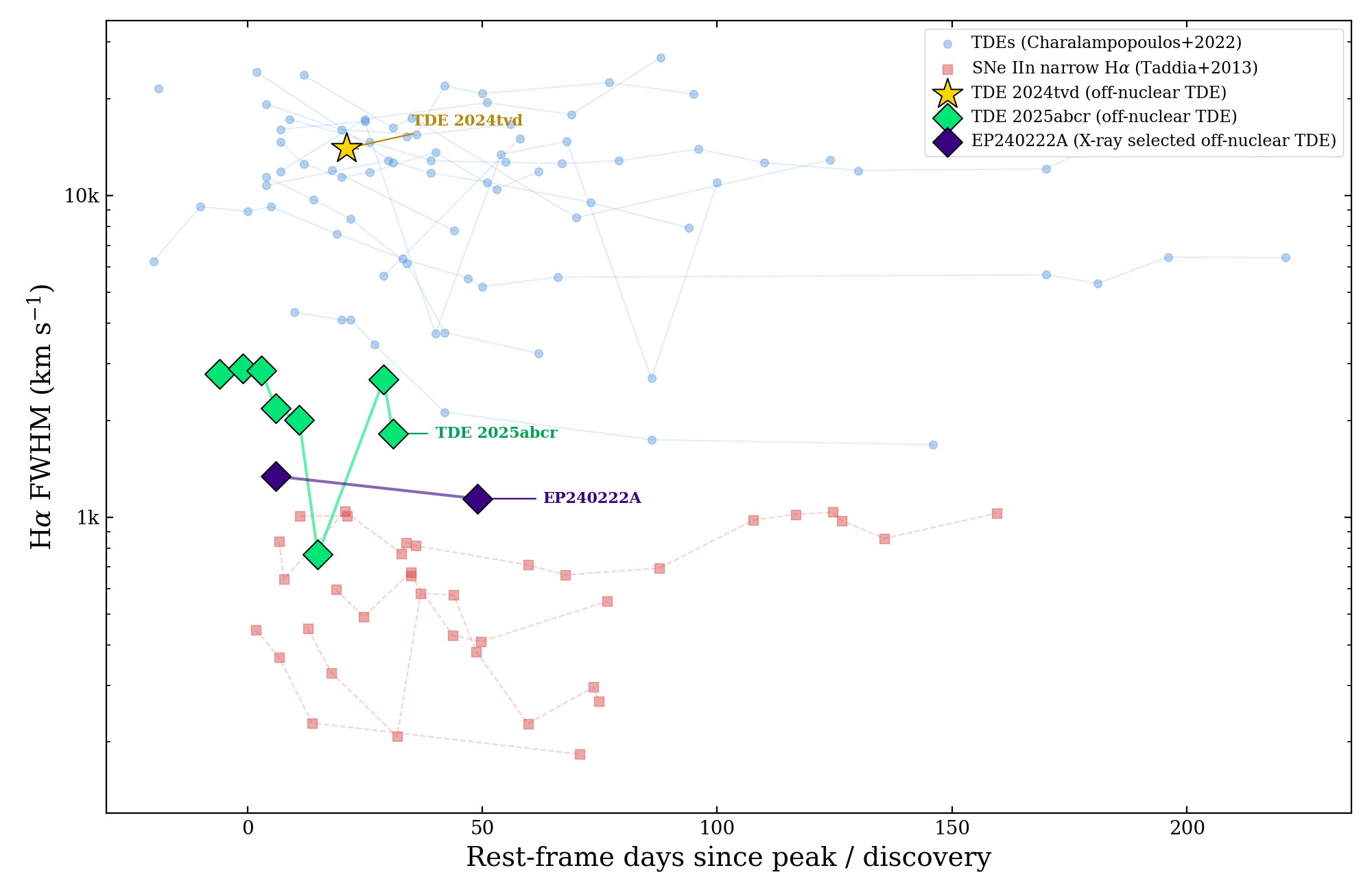}
    \caption{Comparison of H$\alpha$ FWHM versus phase for \name, TDE 2024tvd \citep{yao_25}, EP240222A \citep{ep240222a}, and the nuclear TDE sample of \citet{charalampopoulos_22}, measured from single-Lorentzian fits. For context, we also show the narrow components of double-Lorentzian fits to Type IIn supernovae from \citet{tadia_2013} TDEs are shown in days from peak whereas SN are shown in days from discovery as peak epochs were not uniformly available for the SN IIn sample. \name sits on the low end of the nuclear sample but above the narrow emission widths typical of SN IIn.}
    \label{fig:h-alpha-in-context}
\end{figure*}

\begin{figure*}
    \centering
    \includegraphics[width=0.9\linewidth]{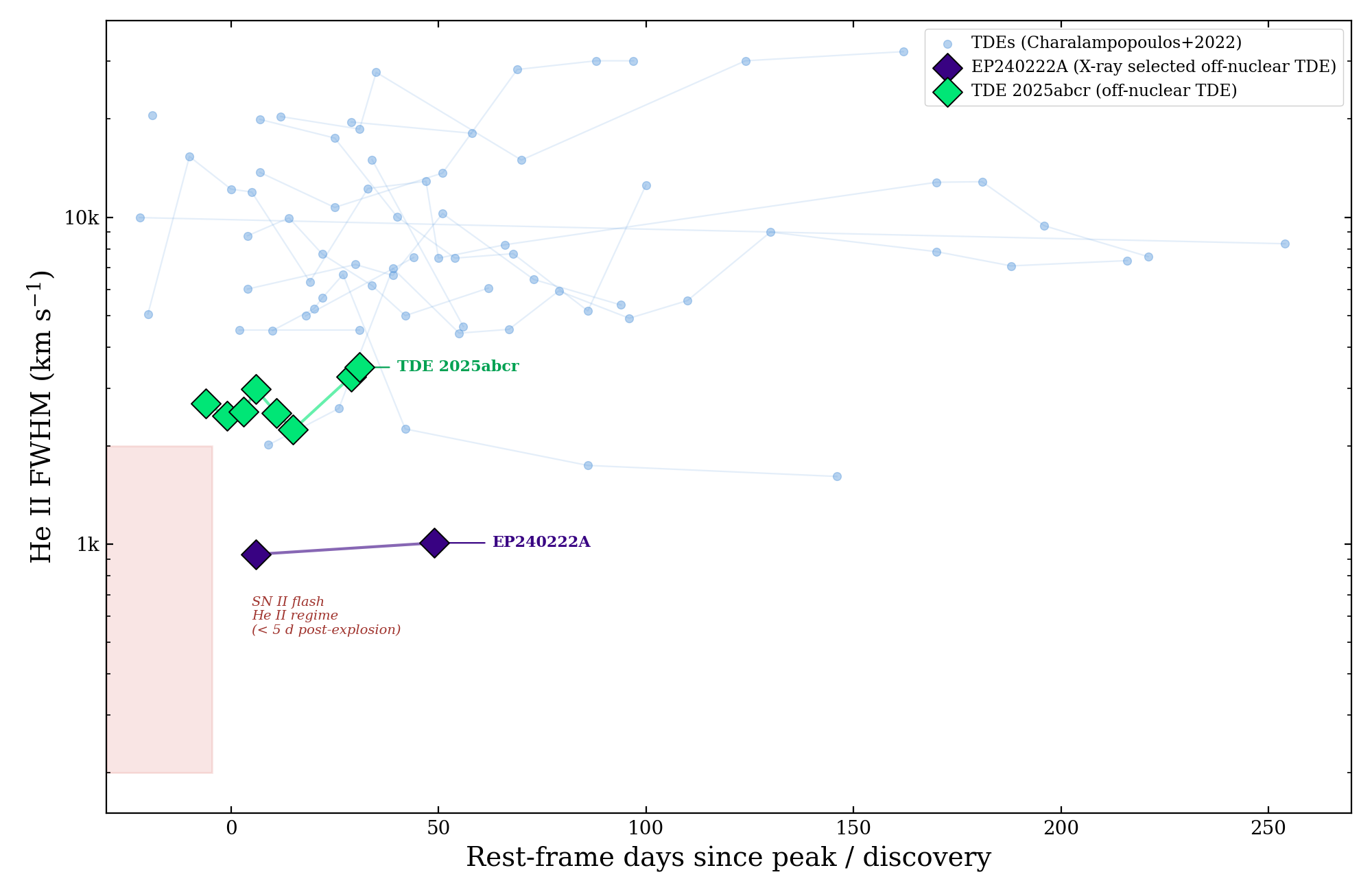}
    \caption{Comparison of He II FWHM vs phase for \name, EP240222A \citep{ep240222a}, and the nuclear TDE sample of \citet{charalampopoulos_22}. \name's He II FWHM was measured using a blended-Lorentzian fit of N III $\lambda$4640 and He II $\lambda$4686 lines. For context, we show the region of this parameter space where flash features occurring in young SN \citep{bruch_2021}. \name sits on the narrow end of the nuclear sample but is broader and longer lasting than those seen in flash features.}
    \label{fig:he-II-in-context}
\end{figure*}

\clearpage

\bibliography{sample701}{}
\bibliographystyle{aasjournalv7}



\end{document}